\documentclass[a4paper,preprint,showpacs,superscriptaddress]{revtex4}
\usepackage[version=3]{mhchem}
\usepackage{float,fancyhdr,amsmath,graphicx,subfig,epsfig,color,setspace,url}
\usepackage{appendix}
\begin{document}

\title{Dissociative Electron Attachment to Polyatomic Molecules - V : Formic Acid and n-Propyl Amine}
\author{N. Bhargava Ram}
\email[]{nbhargavaram@tifr.res.in}
\affiliation{Tata Institute of Fundamental Research, Mumbai 400005, India}

\author{E. Krishnakumar}
\email[]{ekkumar@tifr.res.in}
\affiliation{Tata Institute of Fundamental Research,  Mumbai 400005, India}

\begin{abstract}

In this paper, we discuss the dissociative electron attachment process in Formic Acid and Propyl Amine. These are molecules containing more than one functional group and have low symmetry (\ce{C_{s}} group). We measured the kinetic energy and angular distributions of fragment \ce{H-} ions from the resonances observed in these molecules and compared with that in the precursor molecules, namely - Water, Ammonia and Methane. Measurements suggest that the dissociation dynamics in bigger molecules are independent of overall symmetry of the molecule, rather depend only on the local symmetry of functional group and bond orientation factors.   
\end{abstract}

\maketitle

\section{Introduction}
It has been shown that site selective fragmentation of C-H, N-H and O-H bonds in molecules are possible using electron energy as a control parameter \cite{c6vsp1,c6ptasinska,c6vsp2}. This functional group dependent site selective dissociation has enormous implications from the point of view of chemical control. In order to characterize the dissociation dynamics of the negative ion resonant state and understand the similarities/dissimilarities in various molecules with the same functional group, kinetic energy and angular distribution data are needed. Having made measurements on \ce{H-} ions from the dissociation of OH, NH and CH bond in water, ammonia and methane, we now know the dissociation characteristics in these molecular anions. To compare the dissociation dynamics in these precursor molecules with larger molecules containing similar functional groups, we looked at \ce{H-} ions from Formic acid (\ce{HCOOH} - molecule with CH and OH bonds) and Propyl Amine (\ce{CH3CH2CH2NH2} - molecule with CH and NH bonds).

The choice of these molecules is significant considering their role in the formation of biologically relevant molecules facilitated by the formation of highly reactive radicals due to radiation/charged particle interactions, in particular the resonant processes that occur with high cross section. Formic acid has been observed in several astronomical sources/objects, especially as ices. The abundance of ices rather than gaseous HCOOH is attributed to its low photo-stability \cite{c6schutte,c6ehren}. Therefore, it is important that we understand the gas phase photophysical properties/dynamics of HCOOH and subsequent chemical reactions via intermediate radicals/ions producing larger molecules. Similarly, studying DEA dynamics in propyl amine would be useful in understanding resonant interactions in amino acids. We describe here our finding of similarities/dissimilarities in the DEA dynamics vis-a-vis the precursor molecules of the functional groups. 

\section{Formic Acid}

The ground state geometry of neutral formic acid is planar and belongs to the \ce{C_{s}} symmetry group. The electronic configuration of HCOOH, in the usually accepted Hartree-Fock approximation is ...\ce{(6a^{'})^{2} (7a^{'})^{2} (8a^{'})^{2} (1a^{"})^{2} (9a^{'})^{2} (2a^{"})^{2} (10a^{'})^{2} ->  1A^{$\prime$}}. The two lowest unoccupied molecular orbitals are \ce{3a^{"}}, which is a $\pi^{*}$ molecular orbital and \ce{11a^{'}} which is the antibonding in OH analogue of orbital \ce{9a^{'}}. Studies of excited states of \ce{HCOOH} using VUV absorption spectra and photoelectron spectra \cite{c6leach1,c6schwell,c6leach2} show the lowest energy valence transition \ce{(10a^{'})^{2} -> (10a^{'}) (3a^{"})} , \ce{^{1}A^{'} -> ^{1}A^{"}} corresponding to a \ce{n_{O} -> $\pi$^{*}} transition and observed between 5 and 7 eV. The second valence transition, \ce{1 ^{1}A^{'} -> 2 ^{1}A^{'}} corresponds to the removal of the \ce{2a^{"}} electron and experimentally observed close to 8.1 eV. The third valence transition \ce{1 ^{1}A^{'} -> 2 ^{1}A^{"}} is the \ce{9a^{'} -> 3a^{"}} transition whose energy is calculated to be fairly close to the previous transition. Thus, there are three transitions very closed spaced in the 6 to 9 eV region. One of the characteristics observed in these excitations is the lengthening of the OH bond and reduction in their stretching frequency $\nu_{1}$.

\subsection*{Ion yield curve of \ce{H-}/HCOOH}

\begin{figure}[!htbp]
\centering
\includegraphics[width=0.6\columnwidth]{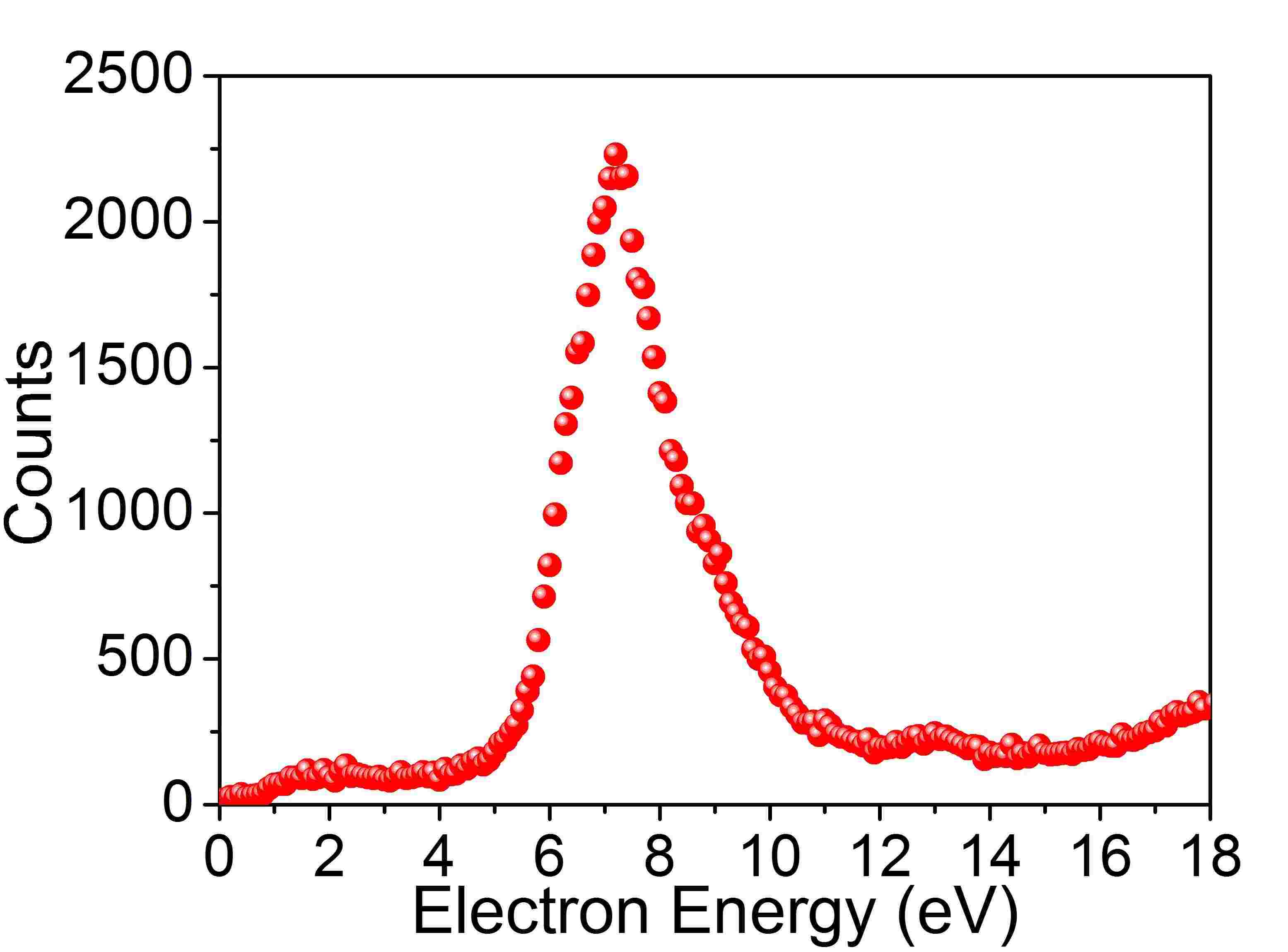}
\caption{Ion yield curve of \ce{H-} from DEA to Formic Acid}
\label{fig6.1}
\end{figure}

The ion yield curve of \ce{H-} from \ce{HCOOH} (Figure \ref{fig6.1}) on dissociative electron attachment shows an asymmetric resonance structure peaking at about 7.5 eV. From the velocity images, we find three distinct scattering patterns - observed at electron energies (i) below 7 eV (ii) between 7 and 8 eV and (iii) above 8 eV. We analyze the \ce{H-}/\ce{HCOOH} velocity images in these three energy intervals and compare them with earlier measurements done using the same setup \cite{c6vsp2} on deuterated acetic acid (\ce{CH3COOD}) and with water and methane (precursor molecules for OH and CH bonds).

\subsection{Results and Discussion}

\subsubsection{Below 7 eV}

\begin{figure}[!htbp]
\centering
\subfloat[\ce{H-}/\ce{HCOOH} at 6.2 eV]{\includegraphics[width=0.3\columnwidth]{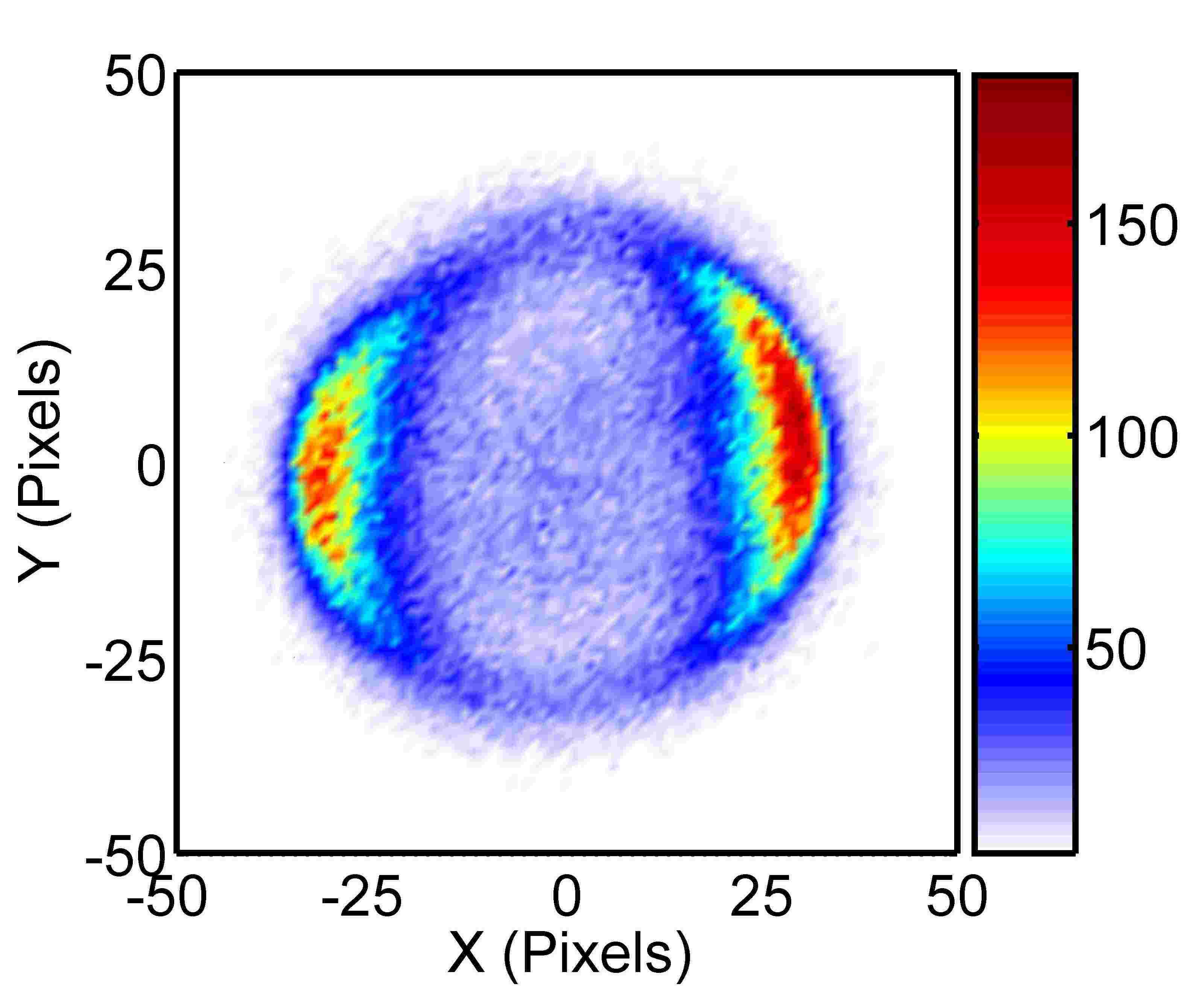}}
\subfloat[\ce{H-}/\ce{HCOOH} at 6.7 eV]{\includegraphics[width=0.3\columnwidth]{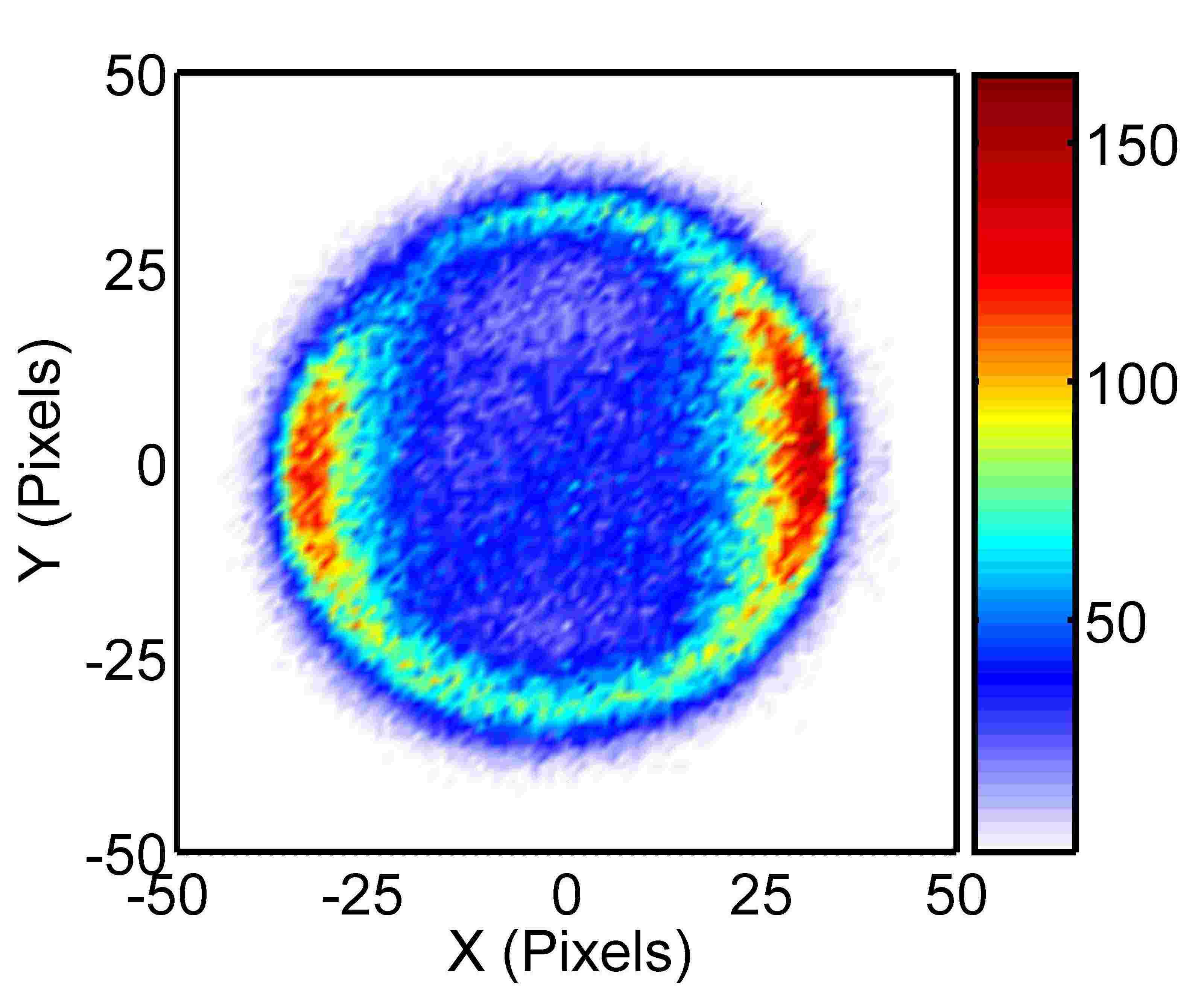}}
\subfloat[\ce{D-}/\ce{CH3COOD} at 6.7 eV]{\includegraphics[width=0.3\columnwidth]{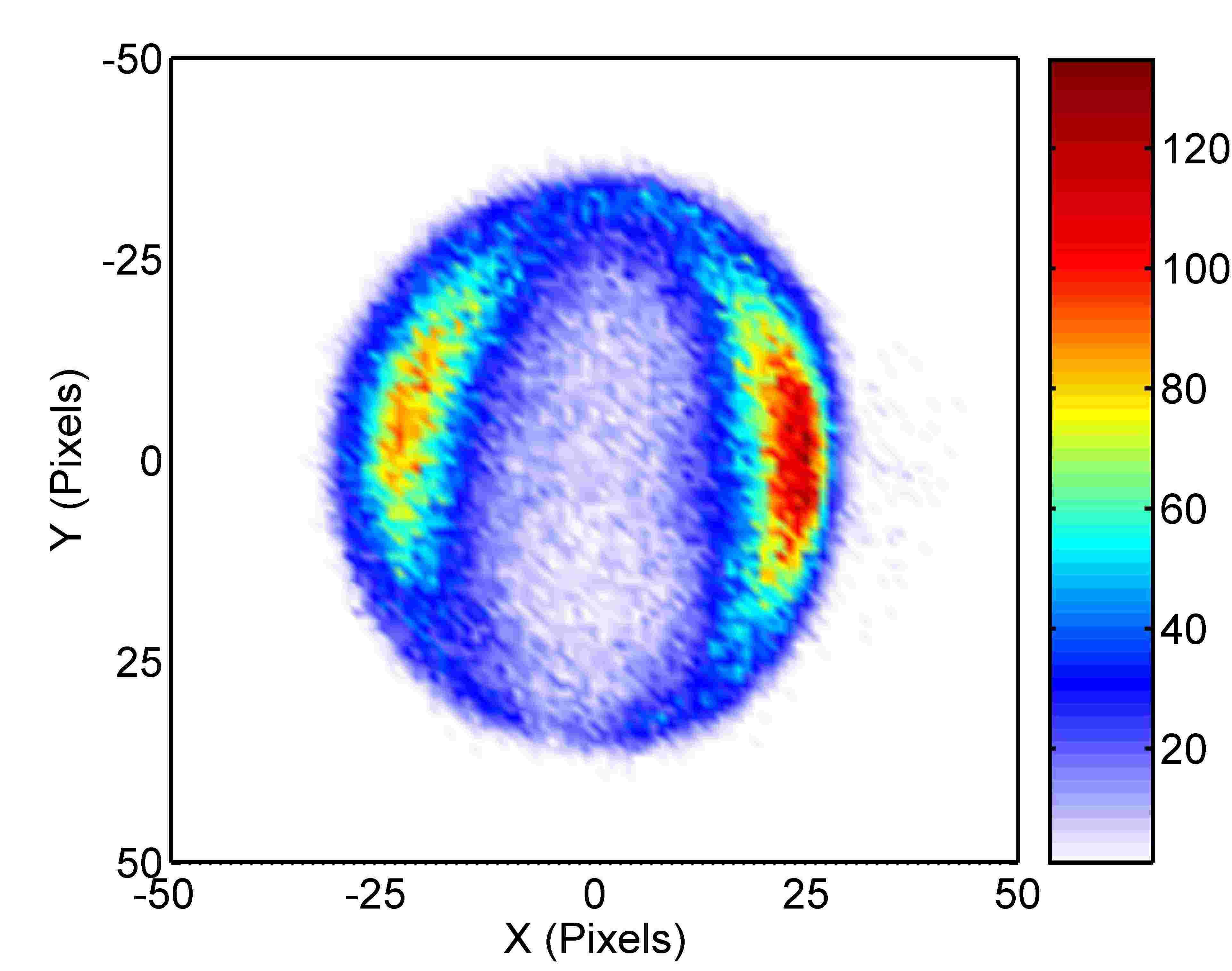}}\\
\subfloat[\ce{H-}/\ce{H2O}]{\includegraphics[width=0.3\columnwidth]{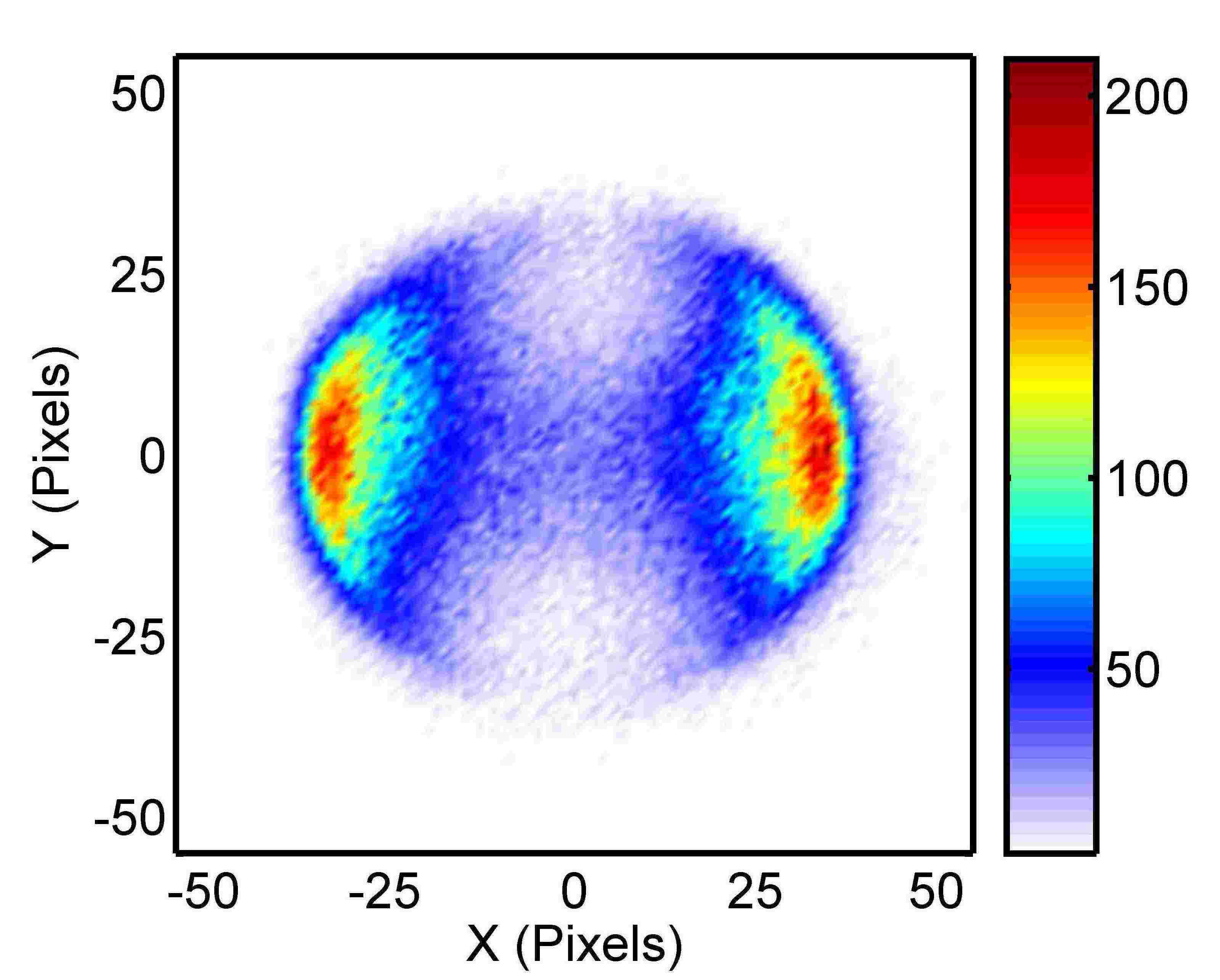}}
\subfloat[KED of \ce{H-}/\ce{HCOOH}]{\includegraphics[height=4cm,width=5cm]{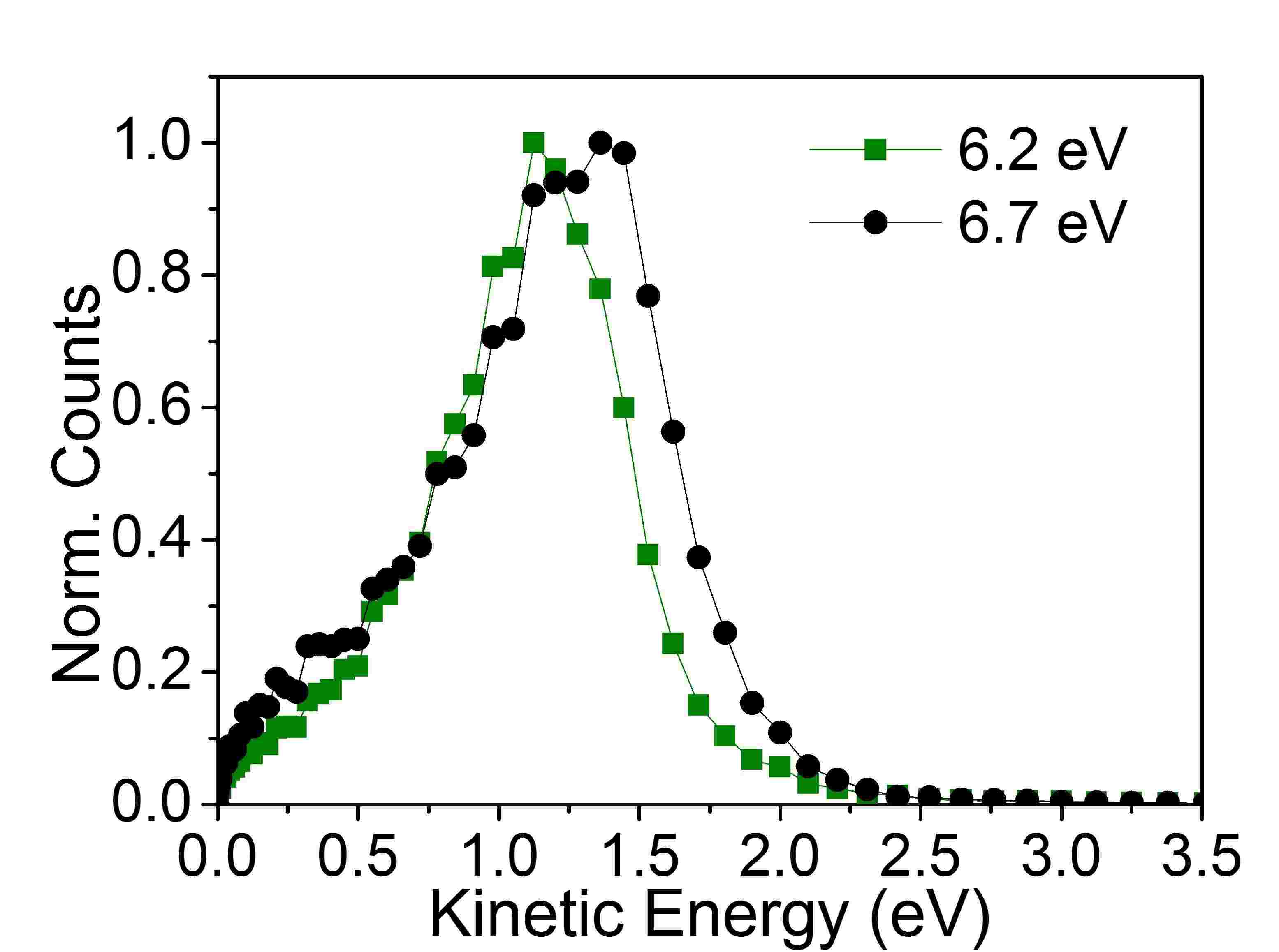}}
\subfloat[Angular distribution plots]{\includegraphics[height=4cm,width=5cm]{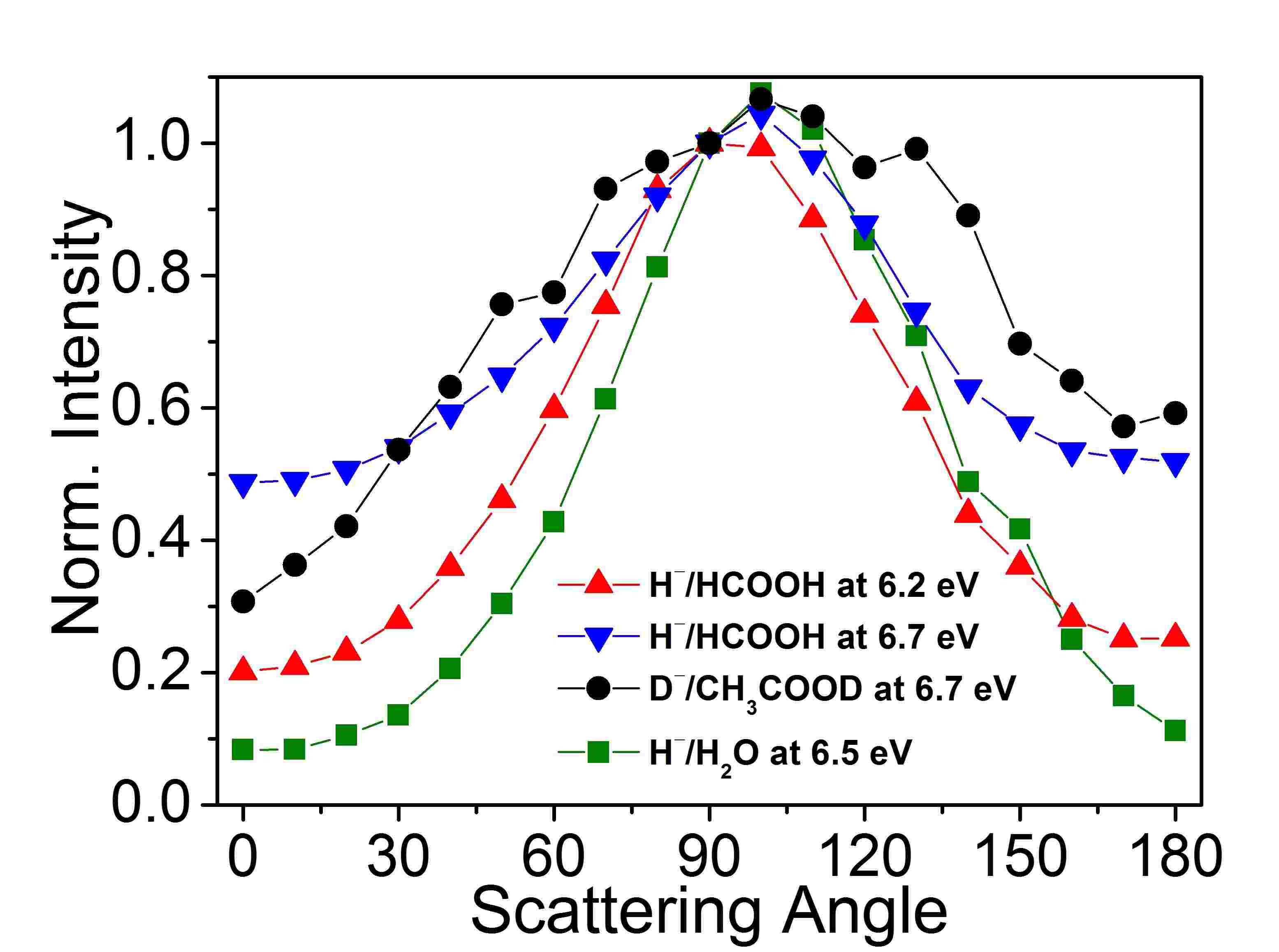}}
\caption{Velocity images of \ce{H-} from DEA to Formic acid at (a) 6.2 eV and (b) 6.7 eV. For comparison, velocity images of (c) \ce{D-}/\ce{CH3COOD} and (d) \ce{H-}/\ce{H2O} at similar electron energies are shown. Kinetic energy distribution of \ce{H-}/\ce{HCOOH} is plotted in (e). And the angular distribution curves for the ions in (a), (b), (c) and (d) are plotted in (f).}
\label{fig6.2}
\end{figure}

\ce{H-} ions at electron energies below 7 eV are mostly scattered in perpendicular direction to the electron beam as seen from the velocity images at 6.2 and 6.7 eV respectively (Figures \ref{fig6.2}(a) and (b)). The kinetic energy distribution at these energies is shown in Figure \ref{fig6.2}(e). The maximum KE observed is about 2.0 eV. The observed \ce{H-} ions could emanate from the dissociation of either the CH bond or the OH bond in HCOOH. The bond dissociation energy of H-COOH and HCOO-H bonds are 3.43eV and 4.54 eV respectively. Electron affinity of H atom being 0.75 eV, the appearance energy for the \ce{H- + COOH} and \ce{HCOO + H-} channels are 2.68 eV and 3.75 eV respectively. Correspondingly, the maximum kinetic energy of \ce{H-} ions produced via these channels at incident electron energy of 6.2 eV will be close to 3.4 eV and 2.3 eV respectively. The KE deduced from the velocity images of \ce{H-} / \ce{HCOOH} show a maximum value of about 2 eV which is close to the maximum KE estimated for H- ions from the OH site i.e. 2.3 eV. The broad KE distribution shows internal excitation of the HCOO fragment. Thus, the \ce{H-} ions observed below 7 eV are attributed to the dissociation of the OH bond in Formic acid.

Further, there exists a distinct similarity in the angular distribution pattern and velocity images of the hydride ions from formic acid, acetic acid and water at the resonance situated at 6.5 eV. In an earlier thesis work \cite{c6vsp2}, measurements on deuterated acetic acid (\ce{CH3COOD}) using the same setup showed similarity between the velocity images of \ce{D-}/\ce{CH3COOD} and \ce{H-}/\ce{H2O}. In all these molecules, the \ce{H-} ions (\ce{D-} in the case of \ce{CH3COOD}) are scattered in the perpendicular direction with respect to the electron beam. Figure \ref{fig6.2}(c) and (d) shows the images of \ce{D-} from \ce{CH3COOD} and \ce{H-} from \ce{H2O} at 6.5 eV and it can be seen that the angular distributions are quite similar (Figure \ref{fig6.2}(f)). Similar results were obtained from measurements on \ce{D-} from \ce{CH3OD} also \cite{c6vsp2}. The similarity in the angular distributions indicates that there is localization of the DEA process to the hydroxyl site and that the OH(OD) bond is perpendicular to the incident electron beam at the instant of attachment giving rise to the angular distribution peaking around $90^{\circ}$. The molecule does not seem to undergo substantial rotation in the dissociation timescale (which is of the order of vibrational time scale) thereby making axial recoil approximation valid at this resonance. Thus, it is the specific orientation of the OH bond undergoing dissociation rather than the whole molecule that determines the observed anisotropy in the angular distribution of \ce{H-} ions. This reaffirms our observation of 'bond orientation specific' DEA process as deduced from the earlier measurements on acetic acid and methanol \cite{c6vsp2} for the resonance process at electron energies close to 6.5 eV. 

Comparing with the VUV absorption spectrum of \ce{HCOOH} \cite{c6leach1}, the resonance process below 7 eV appears to occur via the projectile electron attaching to the \ce{(10a^{'}) (3a^{"}) -> 1 ^{1}A^{"}} excited state giving rise to \ce{(10a^{'}) (3a^{"})^{2}} configuration of the \ce{HCOOH^{-*}} with \ce{^{2}A^{'}} symmetry.

\subsubsection{Between 7 and 8 eV}

\begin{figure}[!ht]
\centering
\subfloat[\ce{H-}/\ce{HCOOH} at 7.2 eV]{\includegraphics[width=0.3\columnwidth]{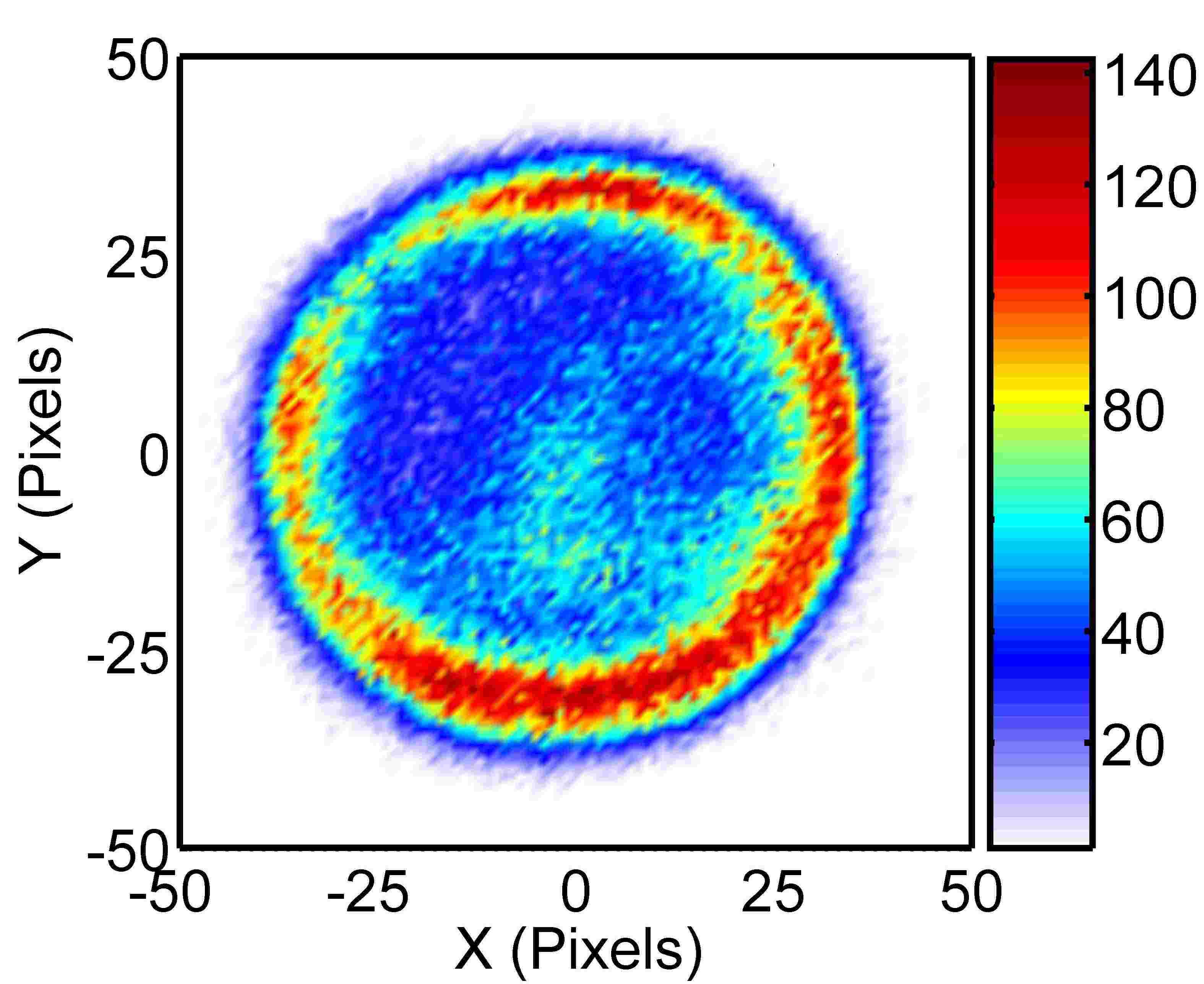}}
\subfloat[\ce{H-}/\ce{HCOOH} at 7.7 eV]{\includegraphics[width=0.3\columnwidth]{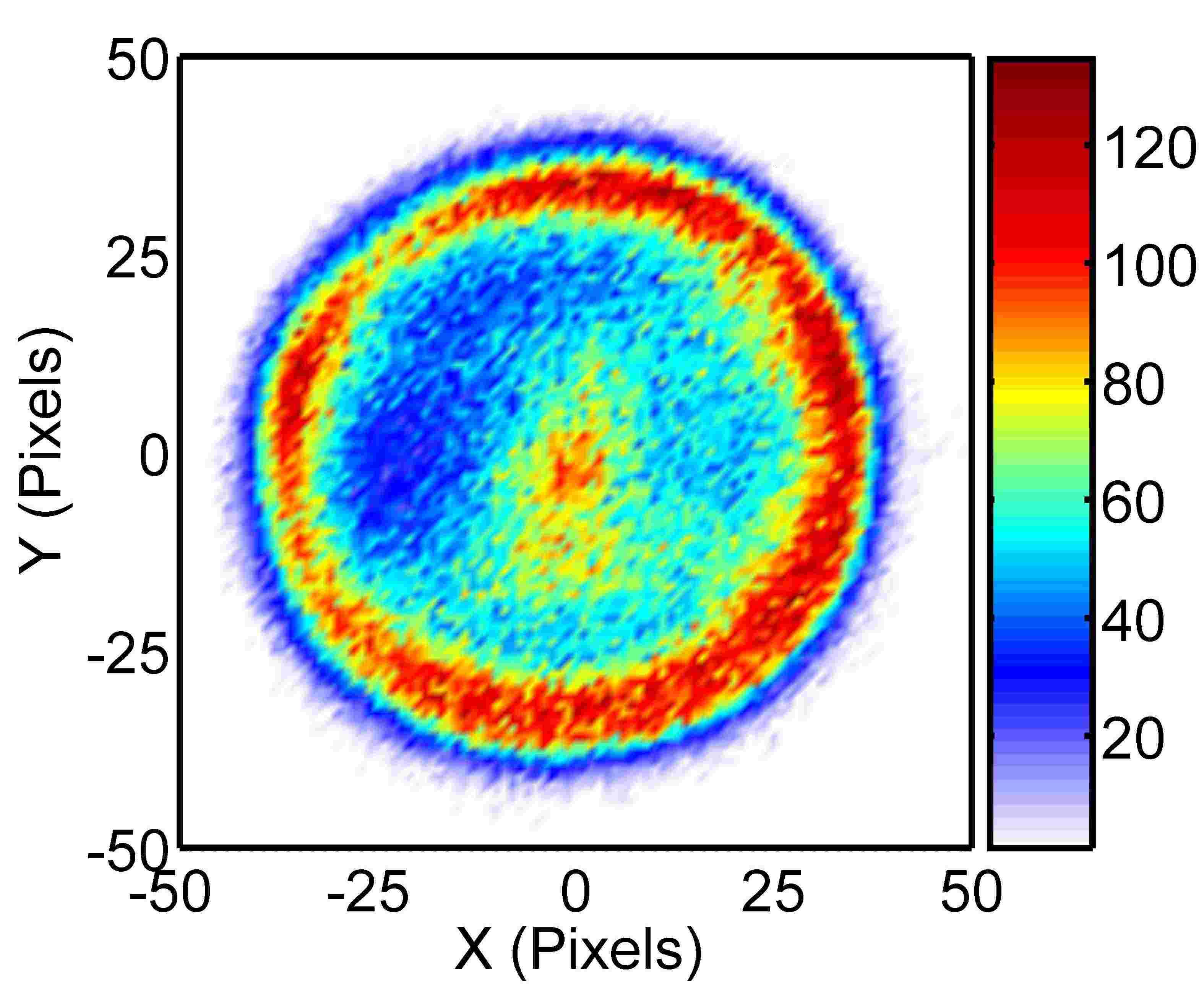}}
\subfloat[\ce{D-}/\ce{CH3COOD} at 7.7 eV]{\includegraphics[width=0.31\columnwidth]{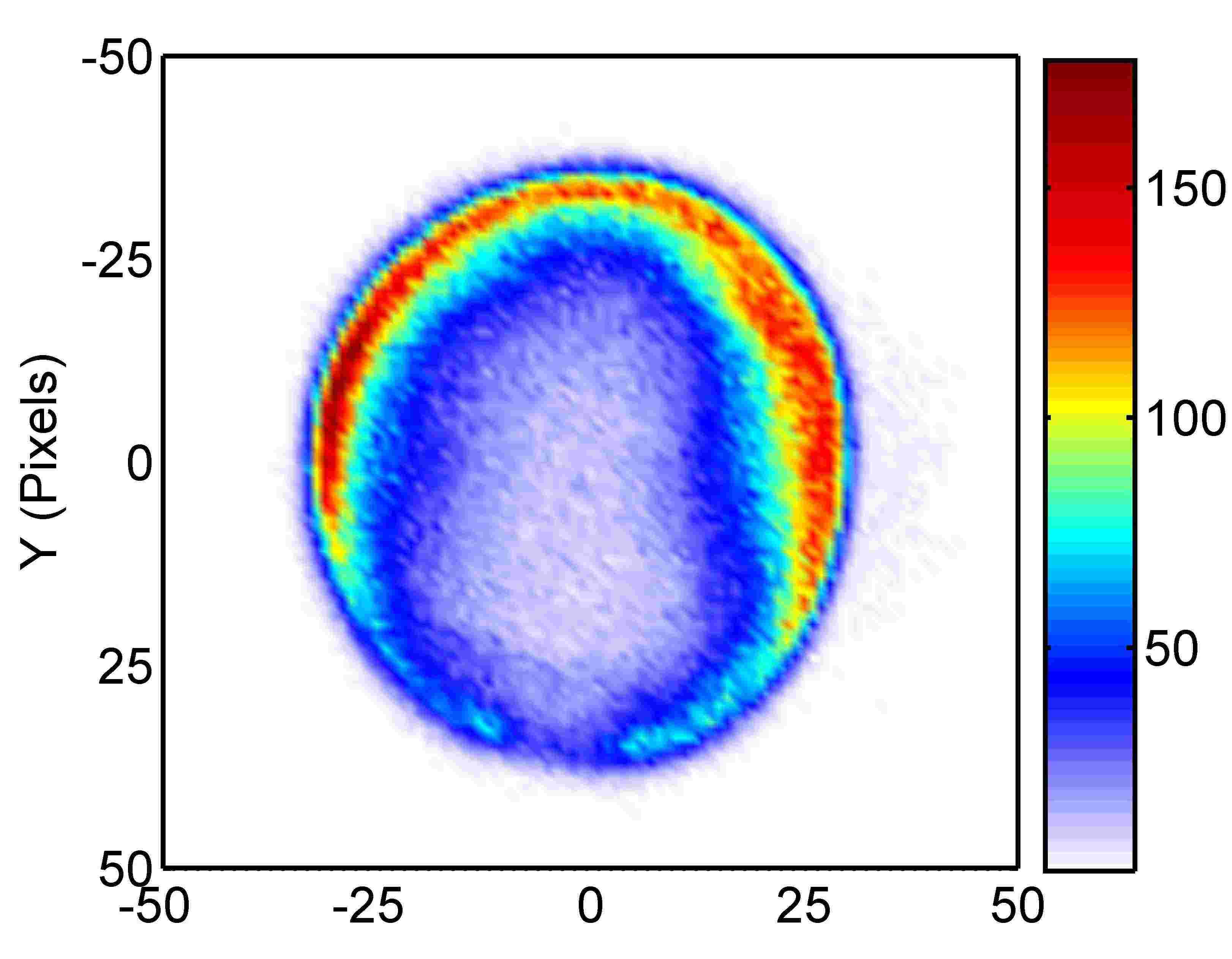}}\\
\subfloat[\ce{H-}/\ce{H2O}]{\includegraphics[width=0.3\columnwidth]{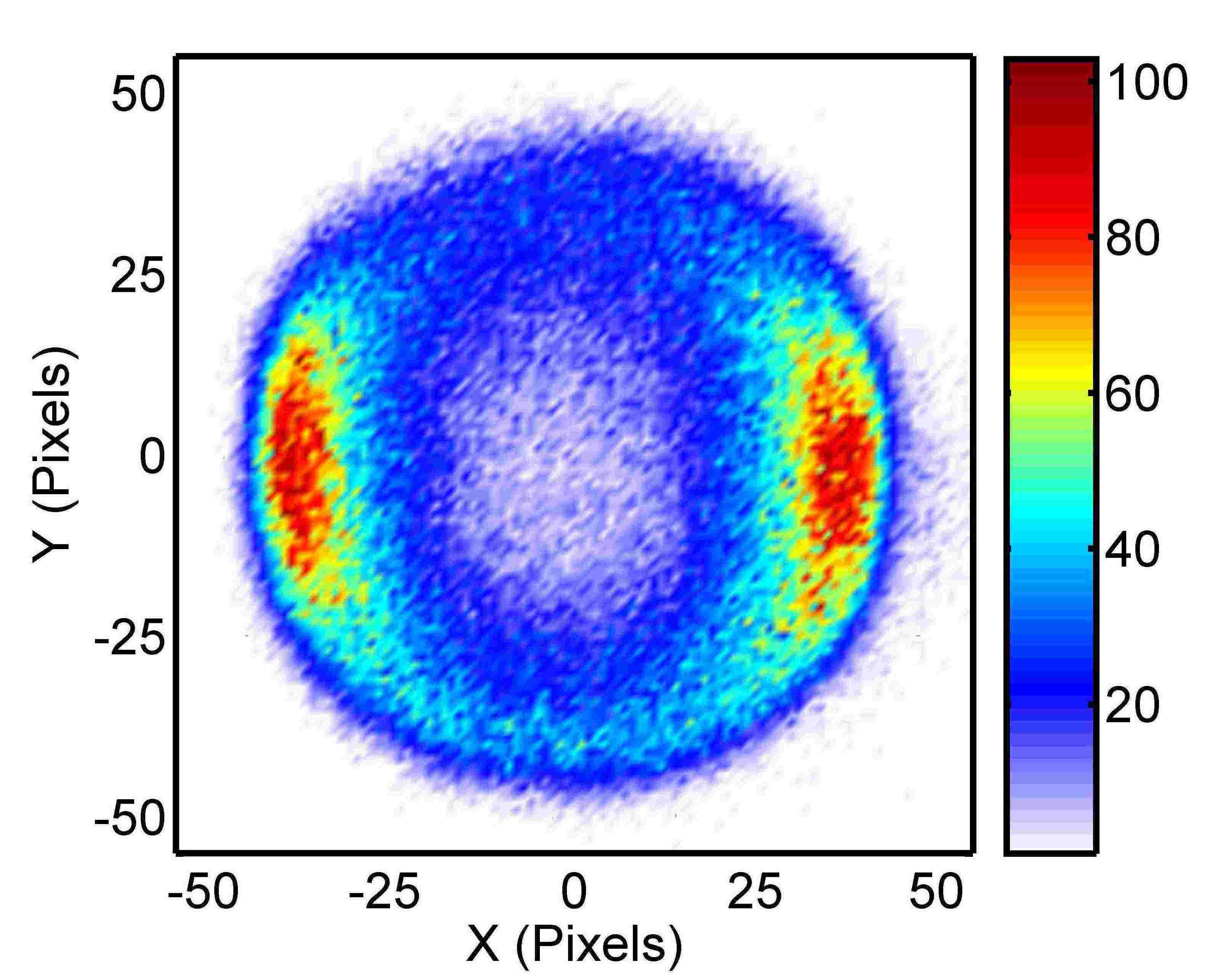}}
\subfloat[KED of \ce{H-} from \ce{HCOOH}]{\includegraphics[width=0.32\columnwidth]{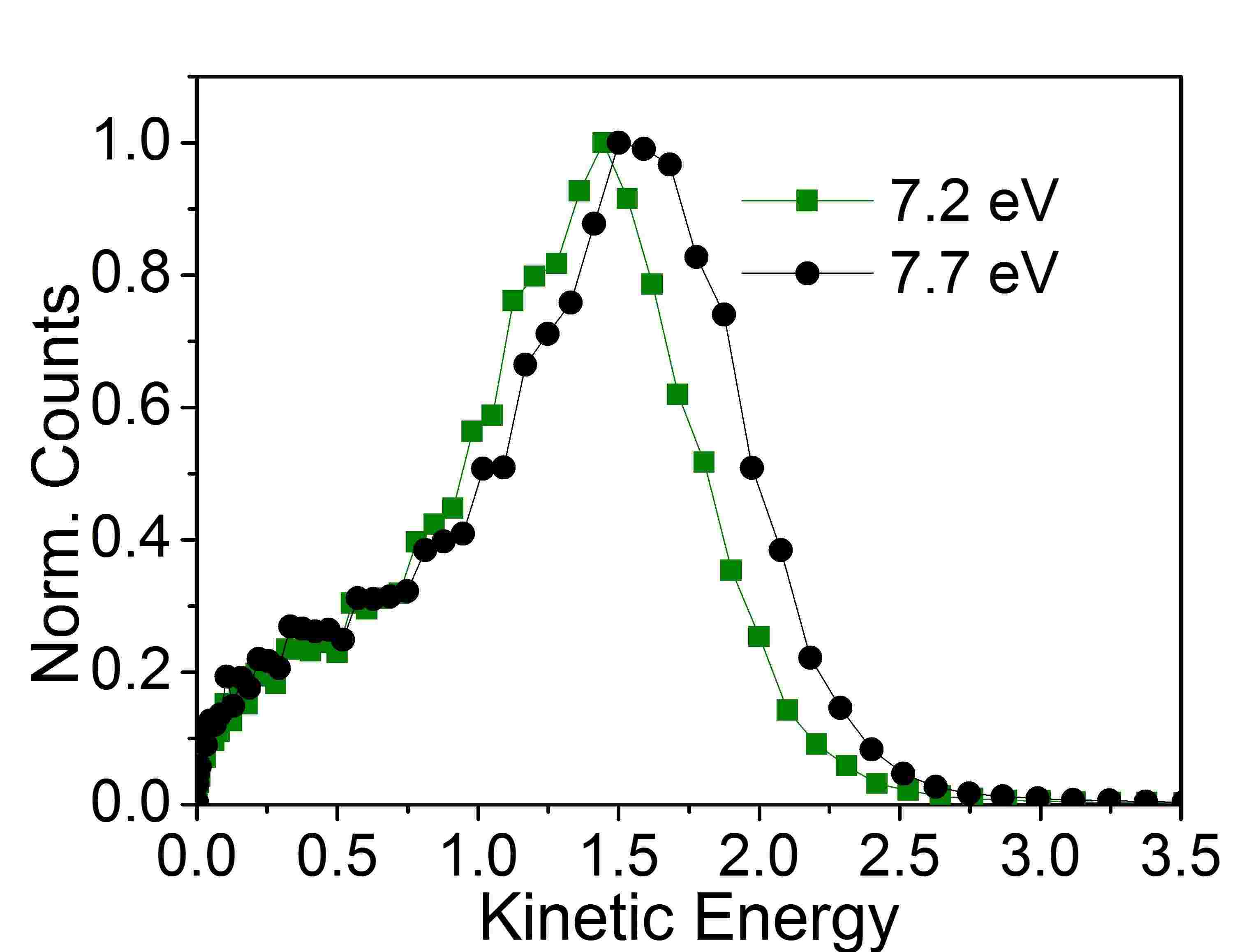}}
\subfloat[Angular distribution plots]{\includegraphics[width=0.32\columnwidth]{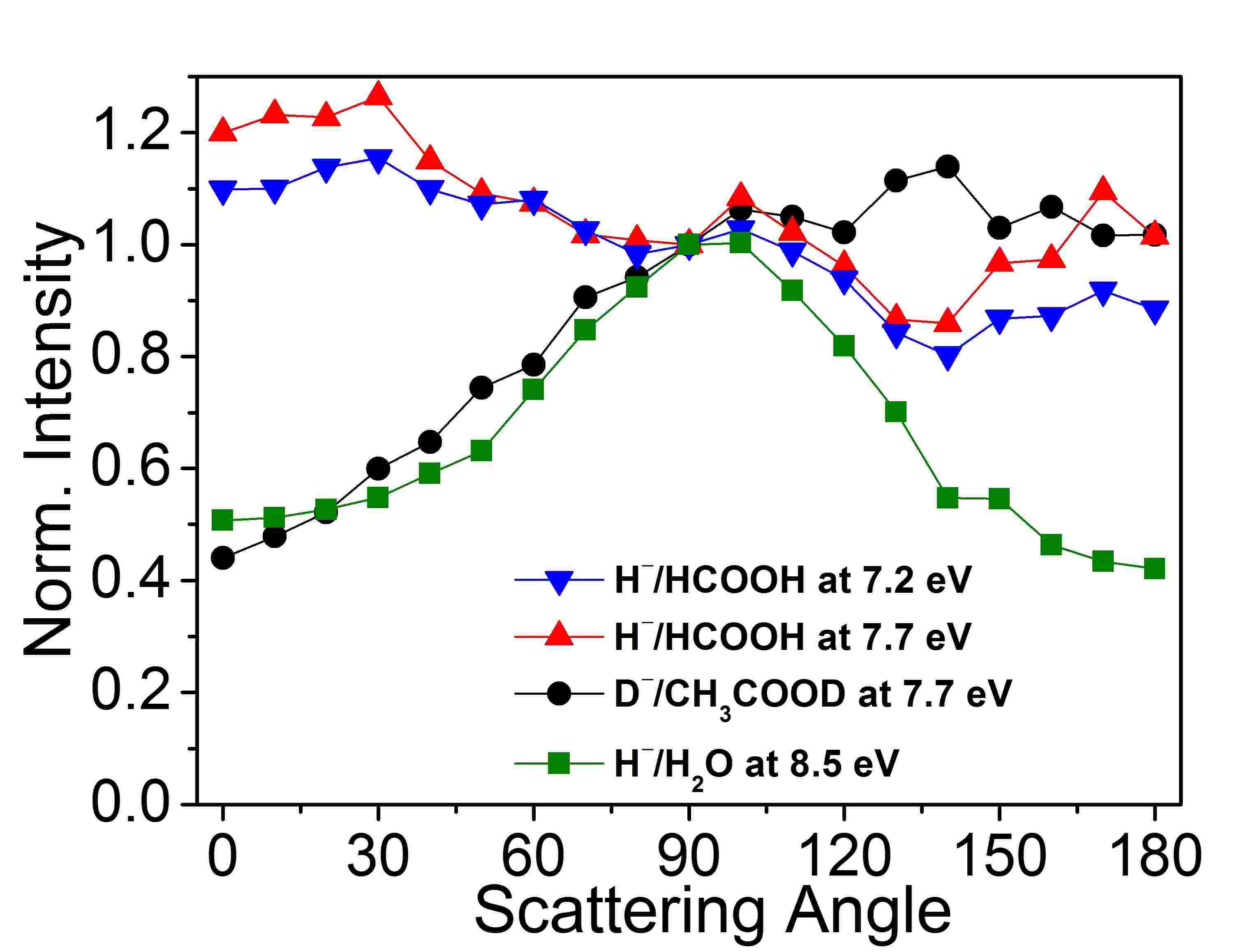}}
\caption{Velocity images of \ce{H-} from DEA to Formic acid at (a) 7.2 eV and (b) 7.7 eV. For comparison, velocity images of (c) \ce{D-}/\ce{CH3COOD} at 7.7 eV and (d) \ce{H-}/\ce{H2O} at 8.5 eV electron energies are shown. Kinetic energy distribution of \ce{H-}/\ce{HCOOH} is plotted in (e). And the angular distribution curves for the ions in (a), (b), (c) and (d) are plotted in (f).}
\label{fig6.3}
\end{figure}

We compare the resonance process in formic acid at these electron energies with the second resonance in water and acetic acid to look for similarities in the dissociation dynamics. The velocity images of \ce{H-} ions at electron energies 7.2 and 7.7 eV show almost an isotropic distribution (see Figures \ref{fig6.3}(a) and (b)). There is a central blob becoming more prominent at 7.7 eV. The maximum KE is seen to be 2.5 eV. For \ce{H-} ions coming from the CH site, the maximum KE expected at 7.2 eV is approximately 4.4 eV and that from the OH site will be 3.3 eV.  As the KE estimated from the dissociation of the OH bond is closer to the observed value of 2.5 eV, it may be the case again that the \ce{H-} ions seen are coming from the OH site with vibrational excitation of the neutral HCOO fragment. 

Figures \ref{fig6.3}(c) and (d) show the velocity images of \ce{D-} from \ce{CH3COOD} at 7.7 eV and \ce{H-} from \ce{H2O} at 8.5 eV and the corresponding angular plots compared with \ce{H-} from \ce{HCOOH} at 7.2 eV and 7.7 eV in Figure \ref{fig6.3}(f). On comparing the angular distribution with \ce{CH3COOD} and \ce{H2O} data, we find that the scattering distributions are not similar to what we see in \ce{H-}/\ce{HCOOH} velocity images as an isotropic distribution. Clearly, they are not consistent. The KE distribution of \ce{H-} from formic acid (Figure \ref{fig6.3}(e)) suggests the dissociation of the OH site. The immediate question is whether the bond orientation specificity properties seen in the case of second resonance in water show up for HCOOH too. However, as the angular distribution of \ce{H-} ions (isotropic distribution) following the dissociation of \ce{HCOOH^{-*}} is seen to be different from that in acetic acid and water (perpendicular distribution), we believe that the isotropic distribution of ions may be a result of a negative ion resonance exclusive to Formic acid. It is possible that this resonance state occurs when a projectile electron with energies in the range 7 to 8 eV attaches to the \ce{2 ^{1}A^{'}} state caused by the excitation of the \ce{2a^{"}} electron. Thus at this resonance, the orientation specificity based dissociation dynamics as seen in the first resonance process (below 7 eV) across the various molecules is absent.

\subsubsection{Above 8 eV}

\begin{figure}[!h]
\centering
\subfloat[\ce{H-}/\ce{HCOOH} at 8.3 eV]{\includegraphics[width=0.3\columnwidth]{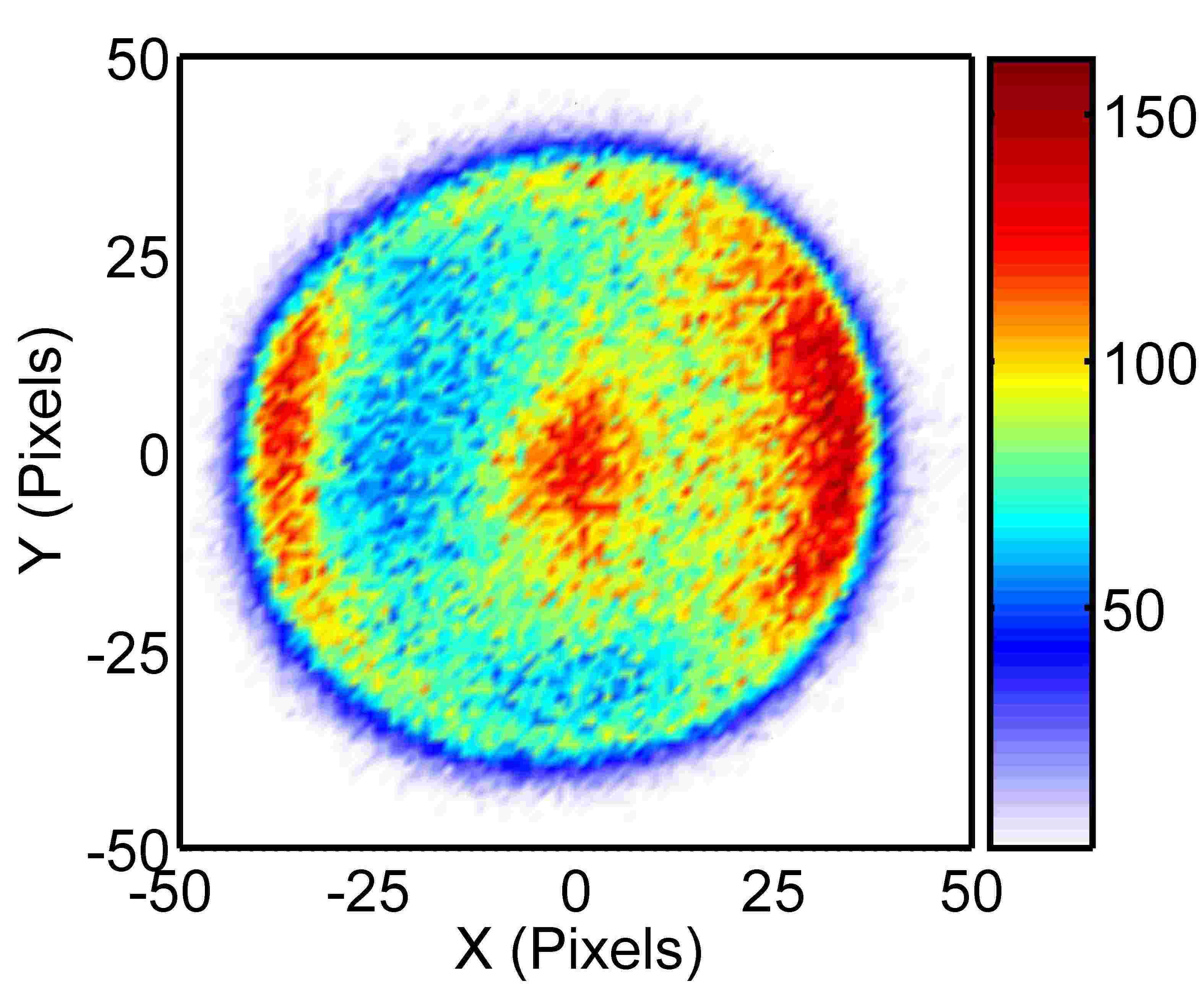}}
\subfloat[\ce{H-}/\ce{HCOOH} at 9.0 eV]{\includegraphics[width=0.3\columnwidth]{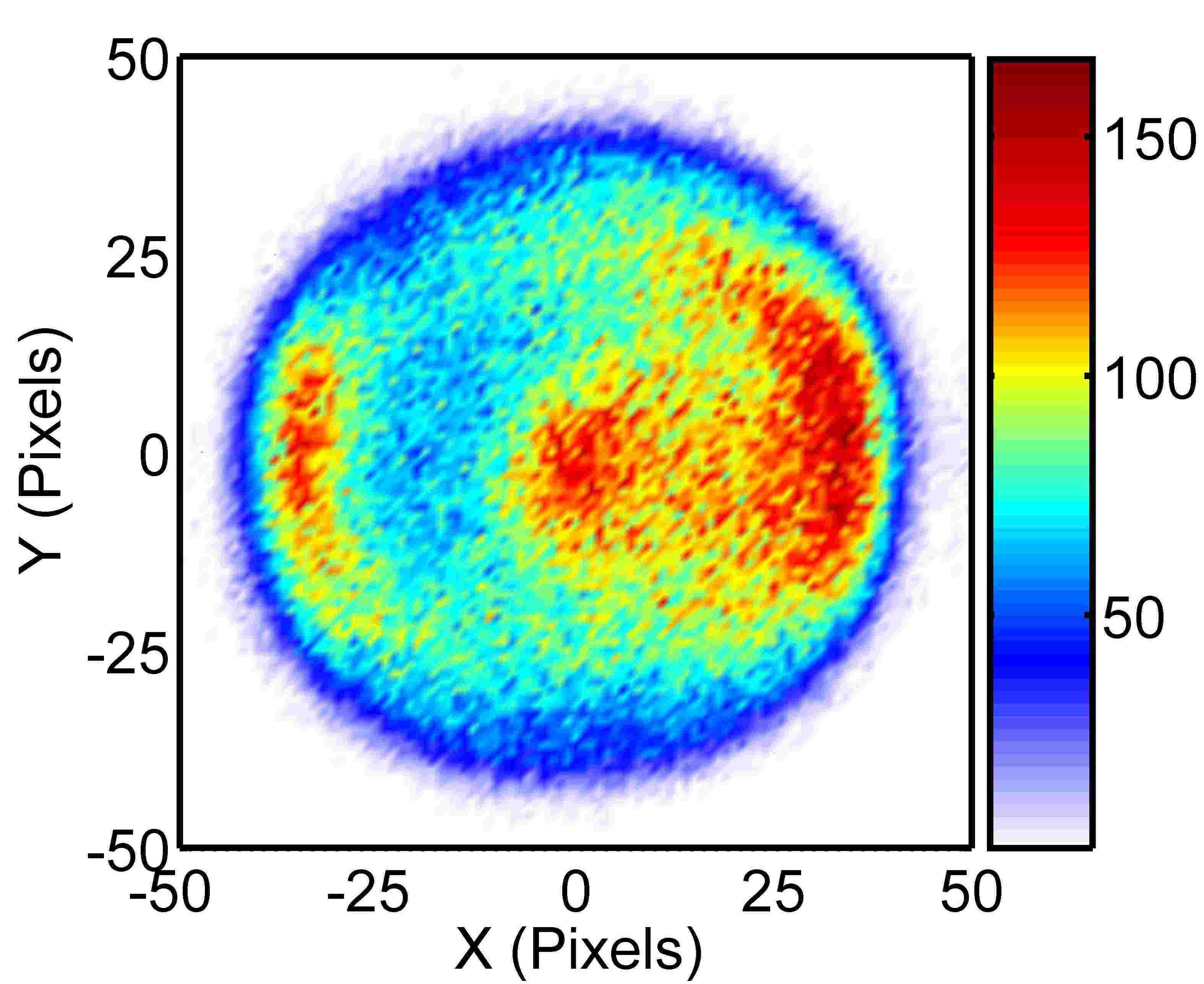}}
\subfloat[\ce{H-}/\ce{HCOOH} at 9.6 eV]{\includegraphics[width=0.3\columnwidth]{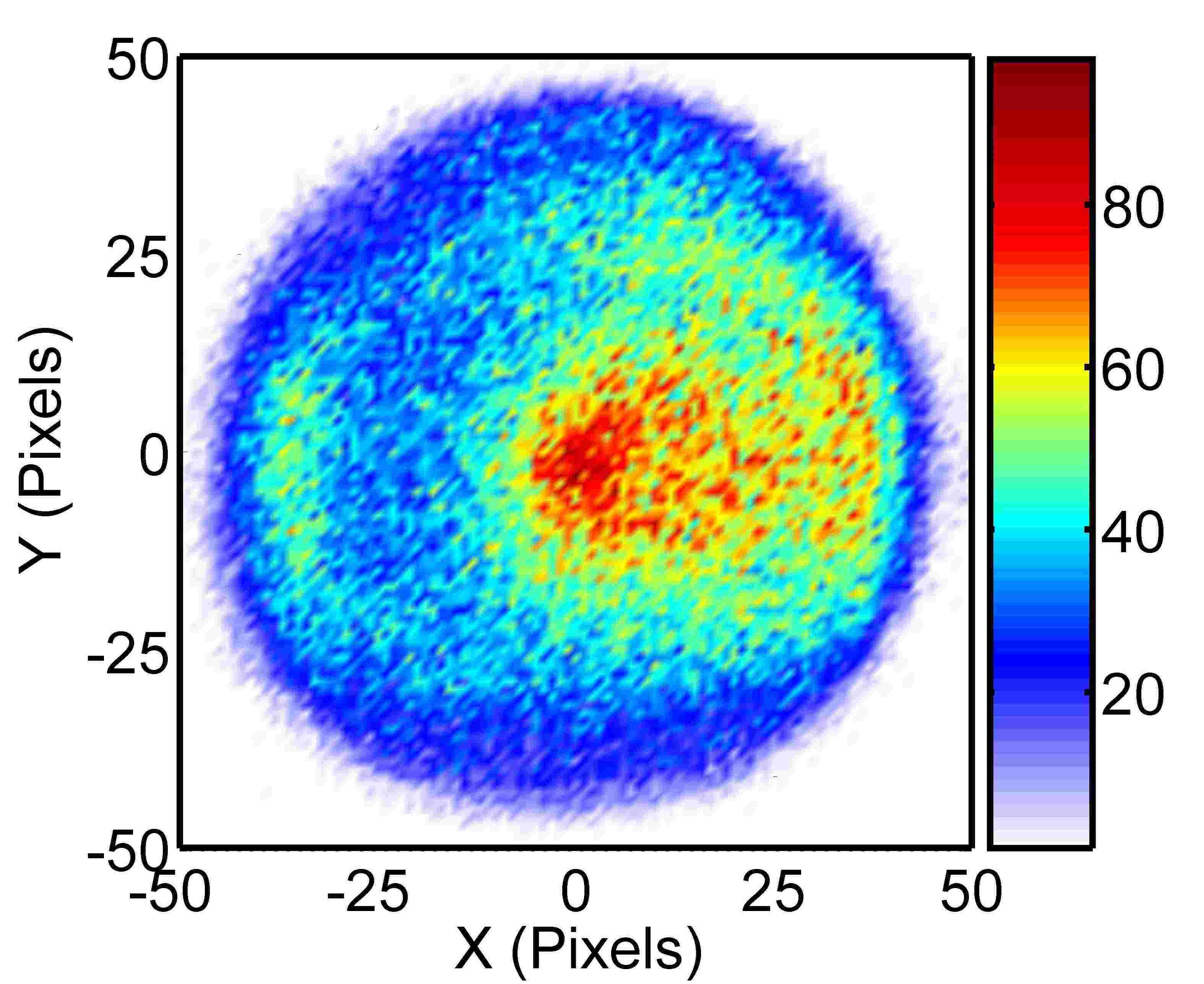}}\\
\subfloat[\ce{D-}/\ce{CH3COOD} at 9.1 eV]{\includegraphics[width=0.3\columnwidth]{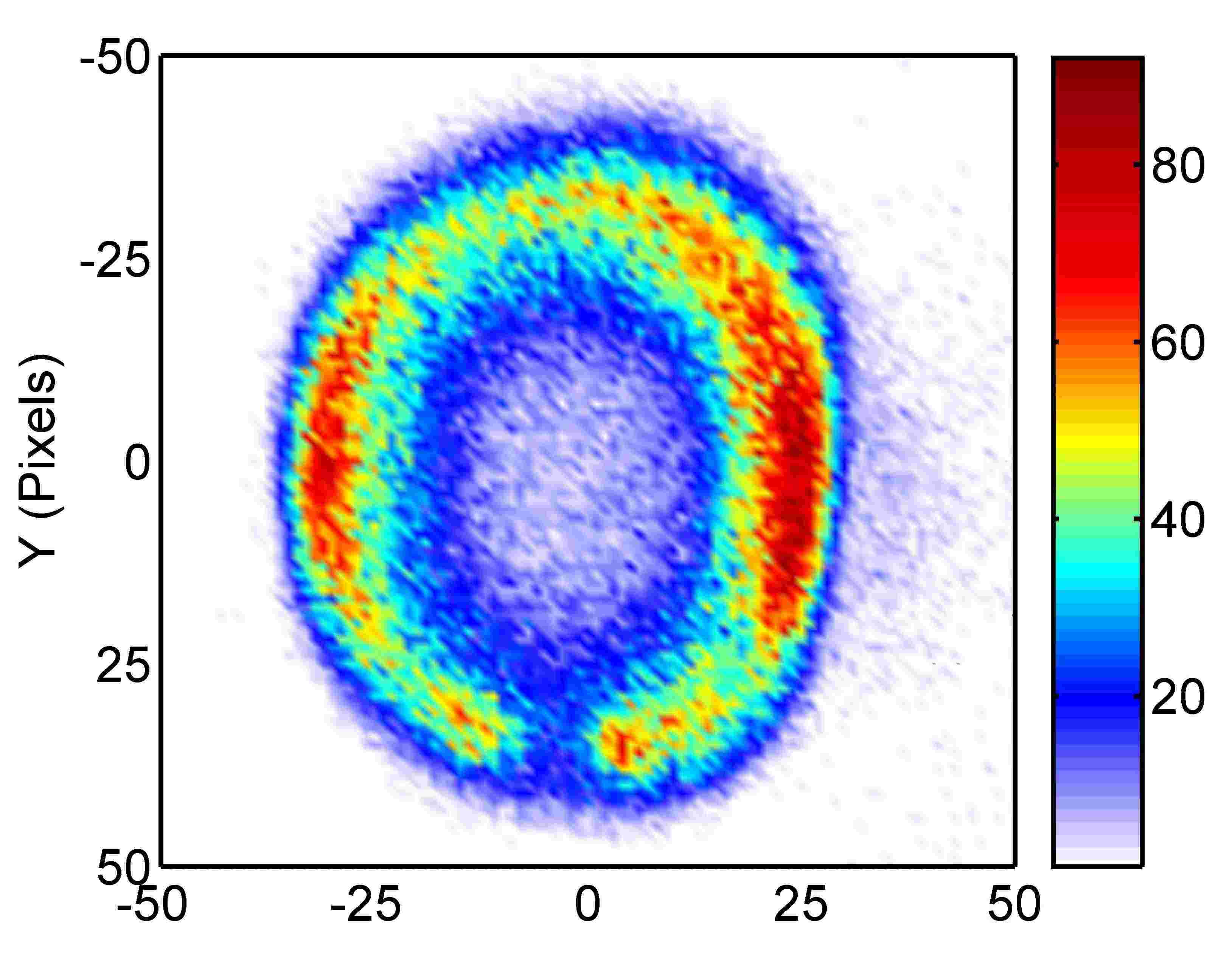}}
\subfloat[\ce{H-}/\ce{CH3COOD} at 9.1 eV]{\includegraphics[width=0.3\columnwidth]{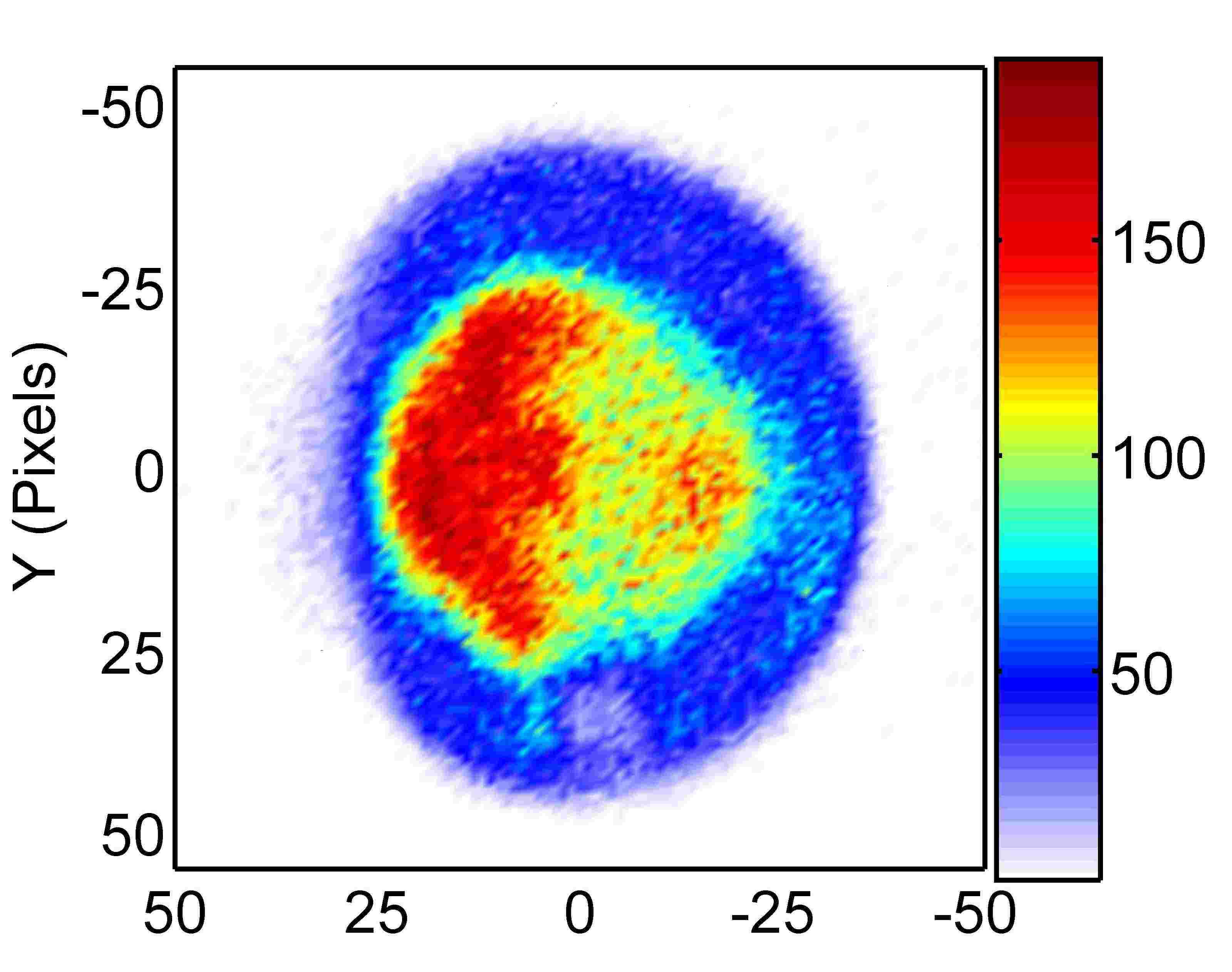}}
\subfloat[\ce{H-}/\ce{CH4} at 10 eV]{\includegraphics[height=4cm,width=5cm]{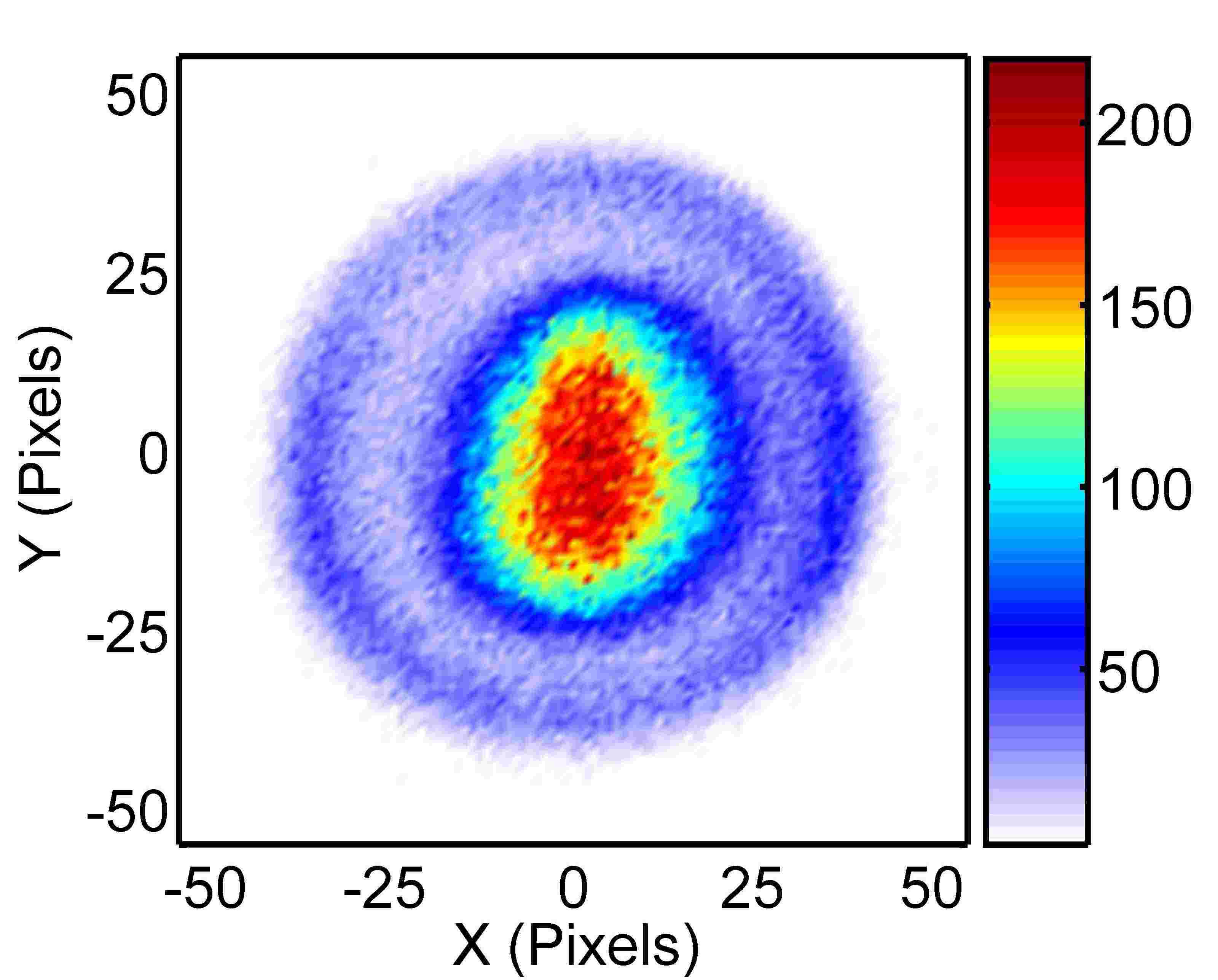}}\\
\subfloat[\ce{H-}/\ce{H2O} at 8.5 eV]{\includegraphics[height=4cm,width=5cm]{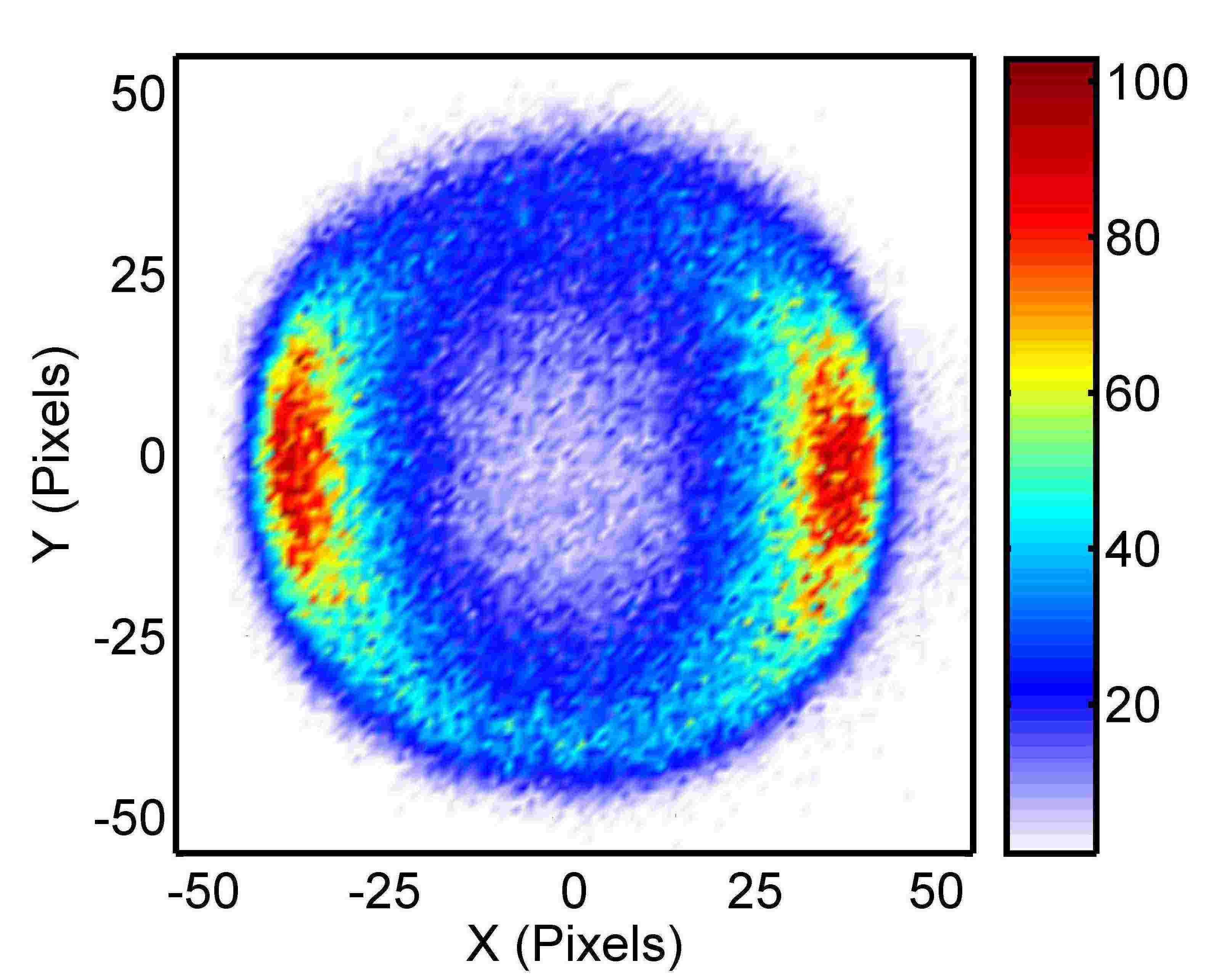}}
\subfloat[KED of \ce{H-} /\ce{HCOOH}]{\includegraphics[height=4cm,width=5cm]{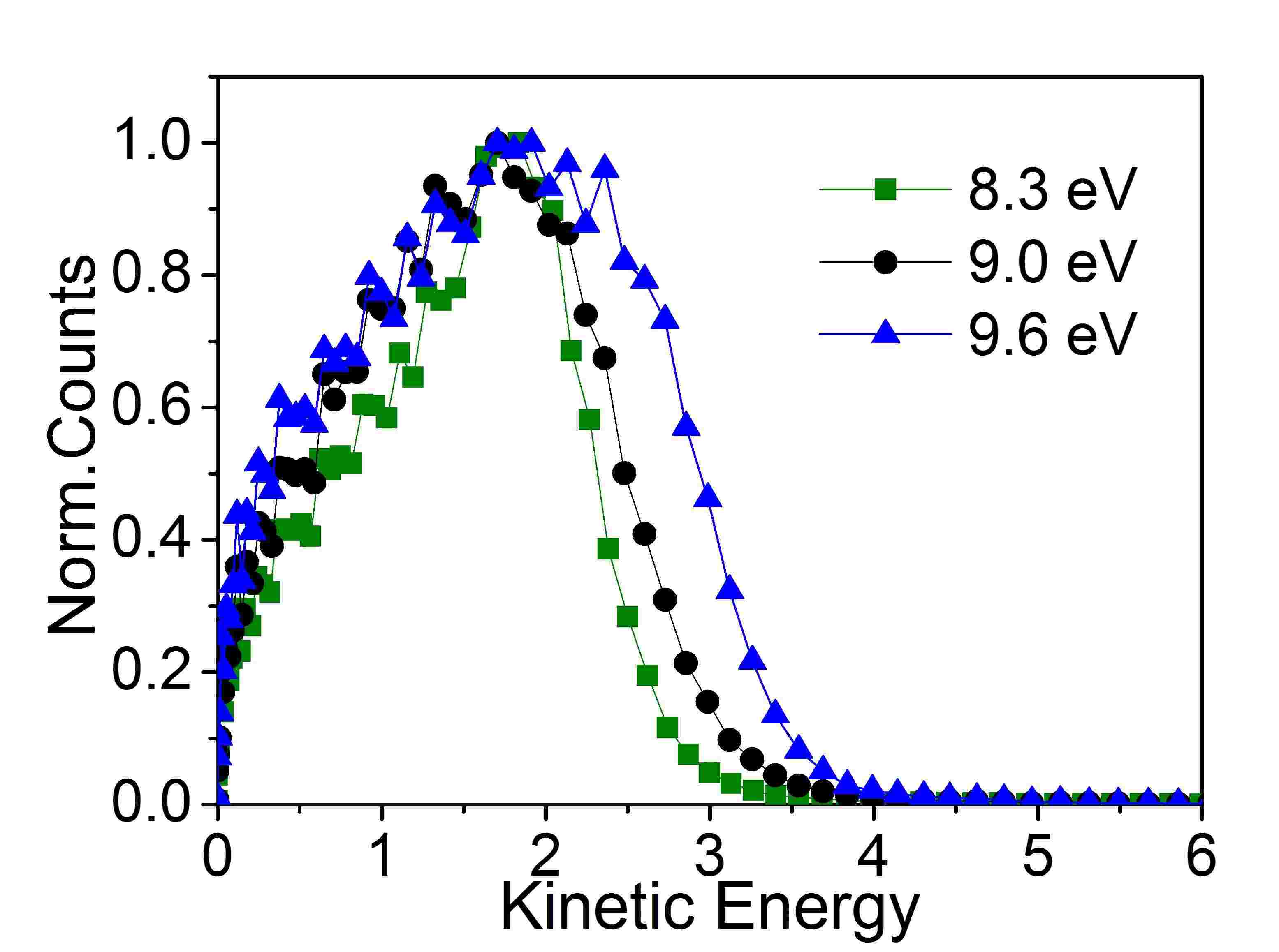}}
\subfloat[Angular distribution plots]{\includegraphics[height=4cm,width=5cm]{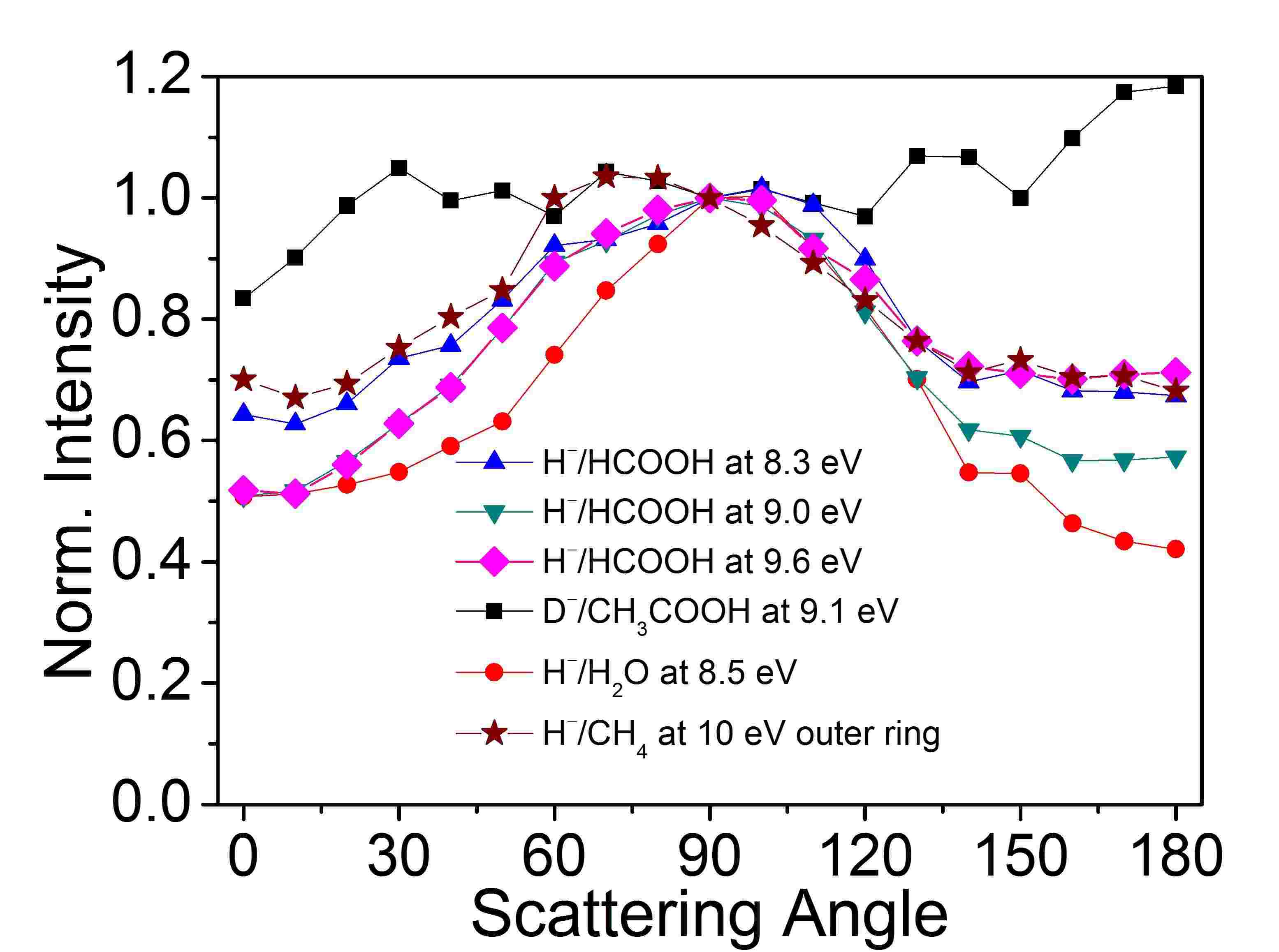}}
\caption{Velocity images of \ce{H-} from DEA to Formic acid at (a) 8.3 eV (b) 9.0 eV and (c) 9.6 eV. For comparison, velocity images of (d) \ce{D-}/\ce{CH3COOD}, (e) \ce{H-}/\ce{CH3COOD}, (f) \ce{H-}/\ce{CH4} and (g) \ce{H-}/\ce{H2O} at similar electron energies are shown. Kinetic energy distribution of \ce{H-}/\ce{HCOOH} is plotted in (h). And the angular distribution curves for the ions in (a), (b), (c),(d) and (g) are plotted in (h).}
\label{fig6.4}
\end{figure}

The velocity images of \ce{H-} ions at electron energies 8.3 eV, 9.0 eV and 9.6 eV are shown in Figure \ref{fig6.4}(a), (b) and (c). There are two distinct structures seen in these images - one is the outer ring with ions scattered in the $90^{\circ}$ direction and the other one is the central blob. The kinetic energy distribution of ions in these three images is plotted in Figure \ref{fig6.4}(g). The maximum kinetic energy is found to be about 3.5 eV. At 9 eV, the \ce{H-} ions from the OH and CH sites are estimated to have a maximum KE of 5.2 and 6.2 eV approximately. Thus, there is a large difference between the measured and the estimated KE values and it is not conclusive as to which bond is being dissociated even while attributing this difference to excitation of various vibration modes. It is also possible that new dissociation channels (other than \ce{H- + COOH} or \ce{HCOO + H-}) resulting via a three-body breakup or in excited electronic state appear. Thus, KE energy distribution cannot single out a particular bond and suggests the possibility of both OH and CH bonds being dissociated.

We compare the \ce{HCOOH} data with measurements done earlier on \ce{CH3COOD}, \ce{CH4} and \ce{H2O} at similar energies and deduce the similarities/dissimilarities. Figures \ref{fig6.4}(d) and (e) show the velocity images of \ce{D-} and \ce{H-} ions from \ce{CH3COOD} and (f) and (g) show the \ce{H-} velocity images from methane at 10 eV and water at 8.5 eV. Comparing the images of these molecules, we see that there are two dissociation limits for the \ce{H-} ions - the outer group and the inner group.

The outer ring seen in the \ce{H-} images from DEA to HCOOH is very similar to the \ce{D-} ions from \ce{CH3COOD} at 9.1 eV with an angular distribution peaking at $90^{\circ}$ direction indicating that the outer ring of \ce{H-} ions emanate from the dissociation of the OH bond. Measurements on \ce{CH3OD} at similar electron energies (close to 9 eV) \cite{c6vsp2} also showed similar behaviour, where the \ce{D-} constituted the outer ring and the \ce{H-} ions formed the inner blob. \ce{H-} ions from \ce{H2O} at 8.5 eV have a predominantly $90^{\circ}$ distribution similar to the outer ring of \ce{H-} ions from Formic acid. However, \ce{H-} ions emanating from Methane via the \ce{H- + CH3} channel also have an angular distribution similar to the outer ring.  Although the measurements in \ce{CH3COOD}, \ce{CH3OD} and \ce{H2O} at similar energies seem to show the dissociation of the hydroxyl site, the DEA data in methane at similar electron energy leaves room to suspect the CH site causing the outer ring distribution of \ce{H-} ions. Measurements on deuterated sample of Formic acid (\ce{HCOOD} or \ce{DCOOH}) are likely to resolve the question of the bond dissociating to give the outer ring. 

Coming to the inner blob seen in the \ce{HCOOH} case, we see that such a blob is seen in the \ce{H-} images from \ce{CH3COOD} and \ce{CH4} both. Such a central blob was also seen in the images of \ce{H-} ions from \ce{CH3OD}. This similarity in the images leads us to conclude that the central blob is due to the \ce{H-} ions coming from the CH site in all these molecules.

 Thus, electron attachment to Formic acid at energies above 8 eV has two distinct dissociation limits with different angular distributions. While the outer distribution (or higher KE) of \ce{H-} ions seem to involve the dissociation of H-COOH or HCOO-H bonds, the inner blob of \ce{H-} ions are attributed to the cleavage of the C-H bond.

\section{Propyl Amine (PA)}

Propyl Amine (PA) - \ce{CH3CH2CH2NH2} is a molecule of \ce{C_{s}} symmetry and contains a propyl group and an amino group. While the three carbon and nitrogen atoms are in a plane, the Hydrogen atoms are out of the plane. Using the GAUSSIAN 03 \cite{c6frisch} program, we find the ground state electronic configuration of neutral PA in Hartree Fock approximation to be ...\ce{(3a^{"})^{2} (11a^{'})^{2} (12a^{'})^{2} (4a^{"})^{2} (13a^{'})^{2} -> 1 ^{1}A^{'}} and the two lowest unoccupied orbitals are \ce{14a^{'}} and \ce{5a^{"}}. The \ce{13a'} molecular orbital is dominantly the \ce{n_{N}} lone pair on the Nitrogen atom of the amino group. We measure the KE and angular distribution of the \ce{H-} ions produced from DEA to PA and look at the dissociation dynamics vis-a-vis measurements on Ammonia and Methane at similar energies. There is no literature available on the VUV absorption or on photoelectron spectra of PA to make correlations/analogies of the resonance features in our measurements with the excited states of the neutral/cation species. 

\begin{figure}[!htbp]
\centering
\includegraphics[width=0.6\columnwidth]{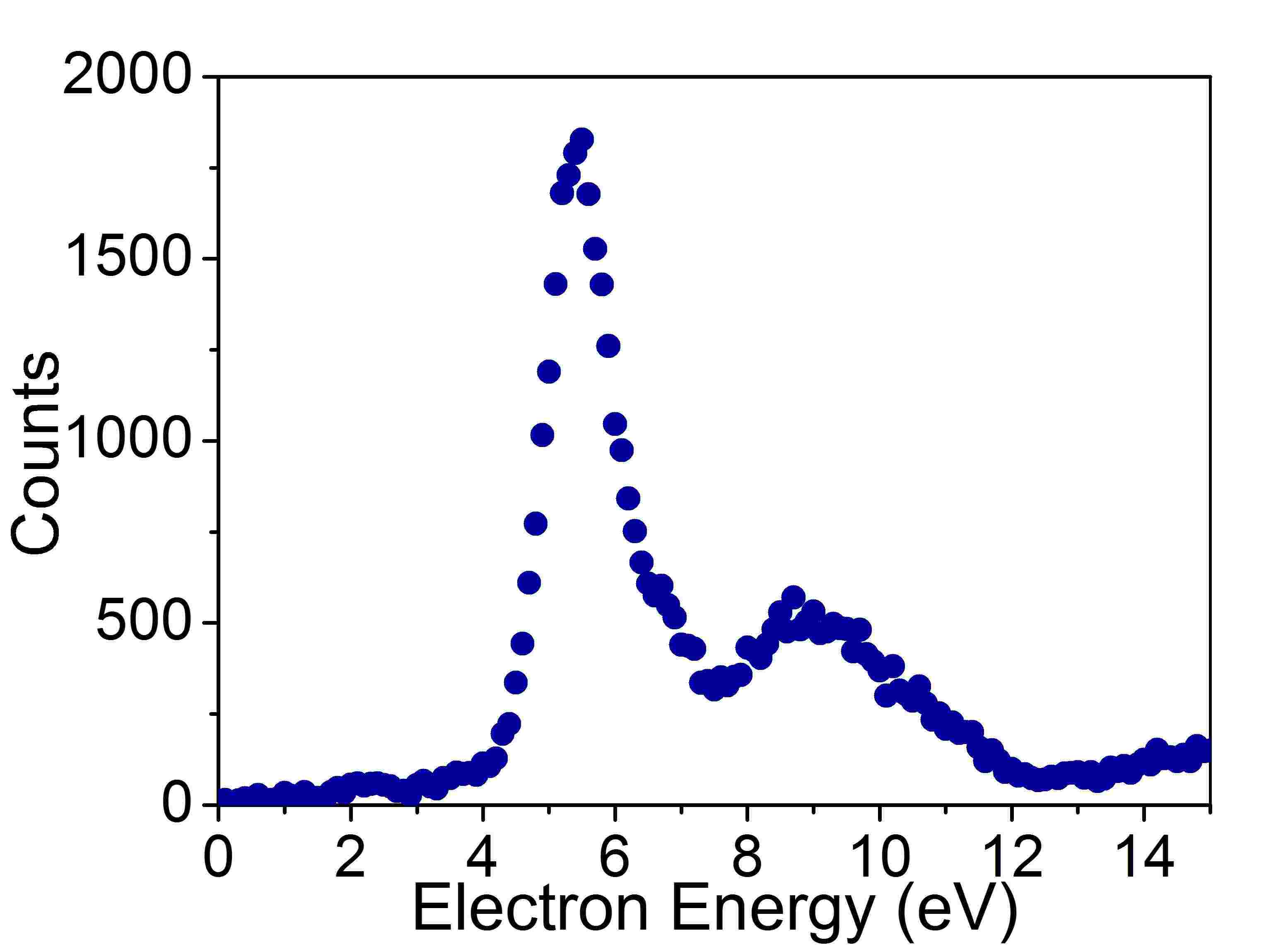}
\caption{Ion yield curve of \ce{H-} from Propyl Amine on electron attachment.}
\label{fig6.5}
\end{figure}

\ce{H-} ion yield on electron attachment to PA shows two structures as in Figure \ref{fig6.5}. The first one is seen with a sharp peak centered at 5.7 eV and the other one is a broad structure in the 8 to 11 eV region. The details are described for the two resonance structures.

\subsection{Results and Discussion}
\subsubsection{First resonance process centered at 5.7 eV}

\begin{figure}[!htbp]
\centering
\subfloat[\ce{H-}/PA at 4.7 eV]{\includegraphics[width=0.3\columnwidth]{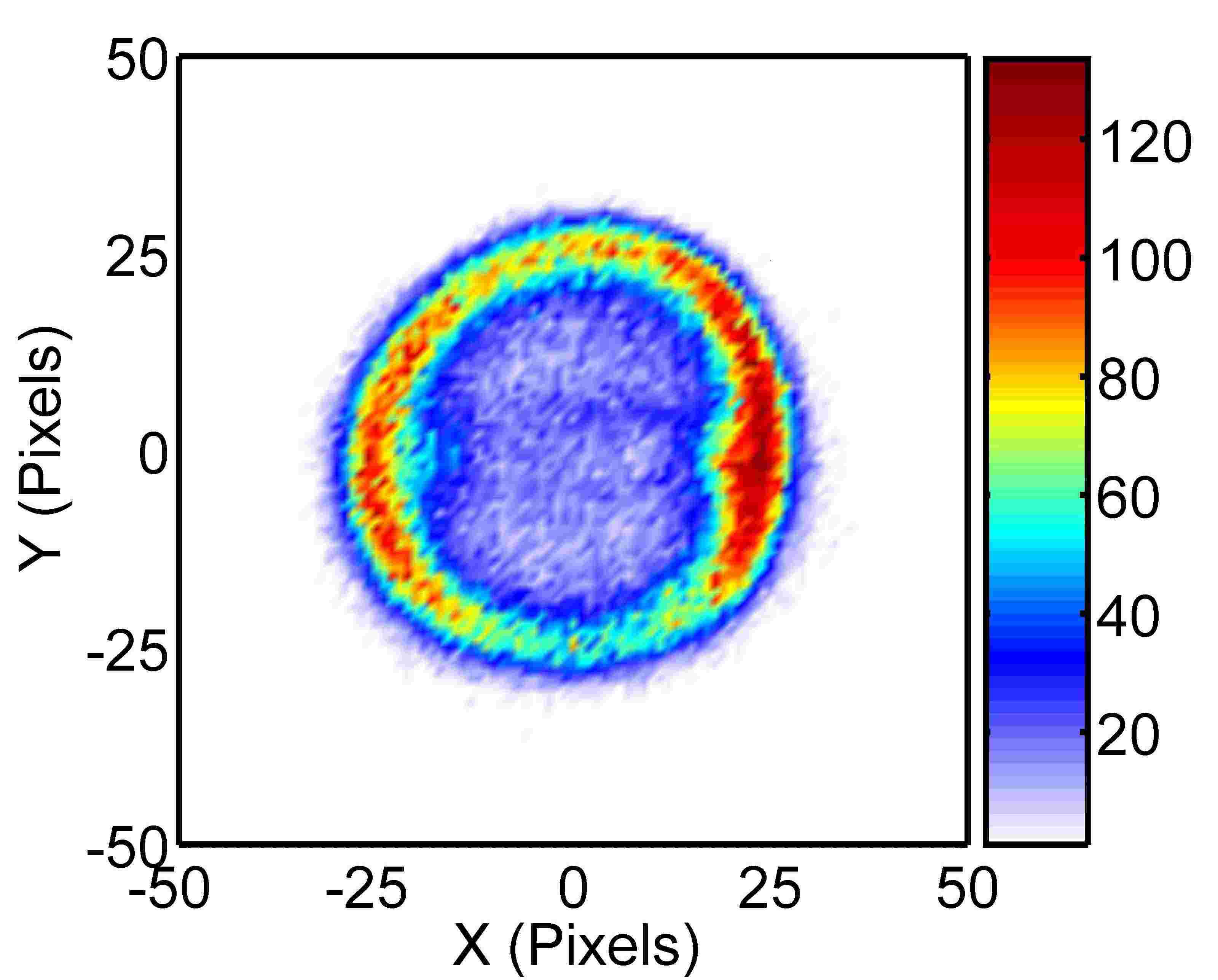}}
\subfloat[\ce{H-}/PA at 5.7 eV]{\includegraphics[width=0.3\columnwidth]{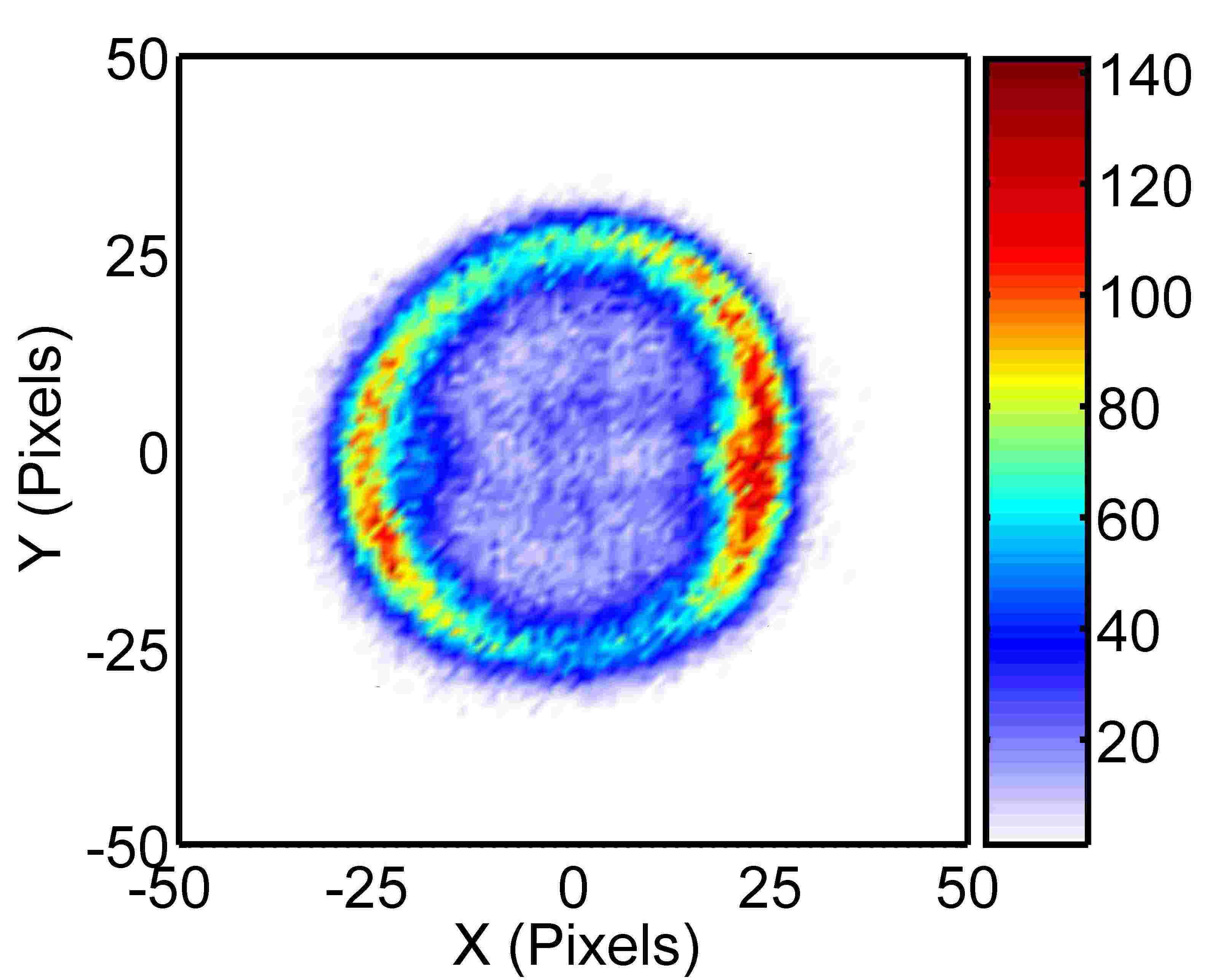}}
\subfloat[\ce{H-}/PA at 6.7 eV]{\includegraphics[width=0.3\columnwidth]{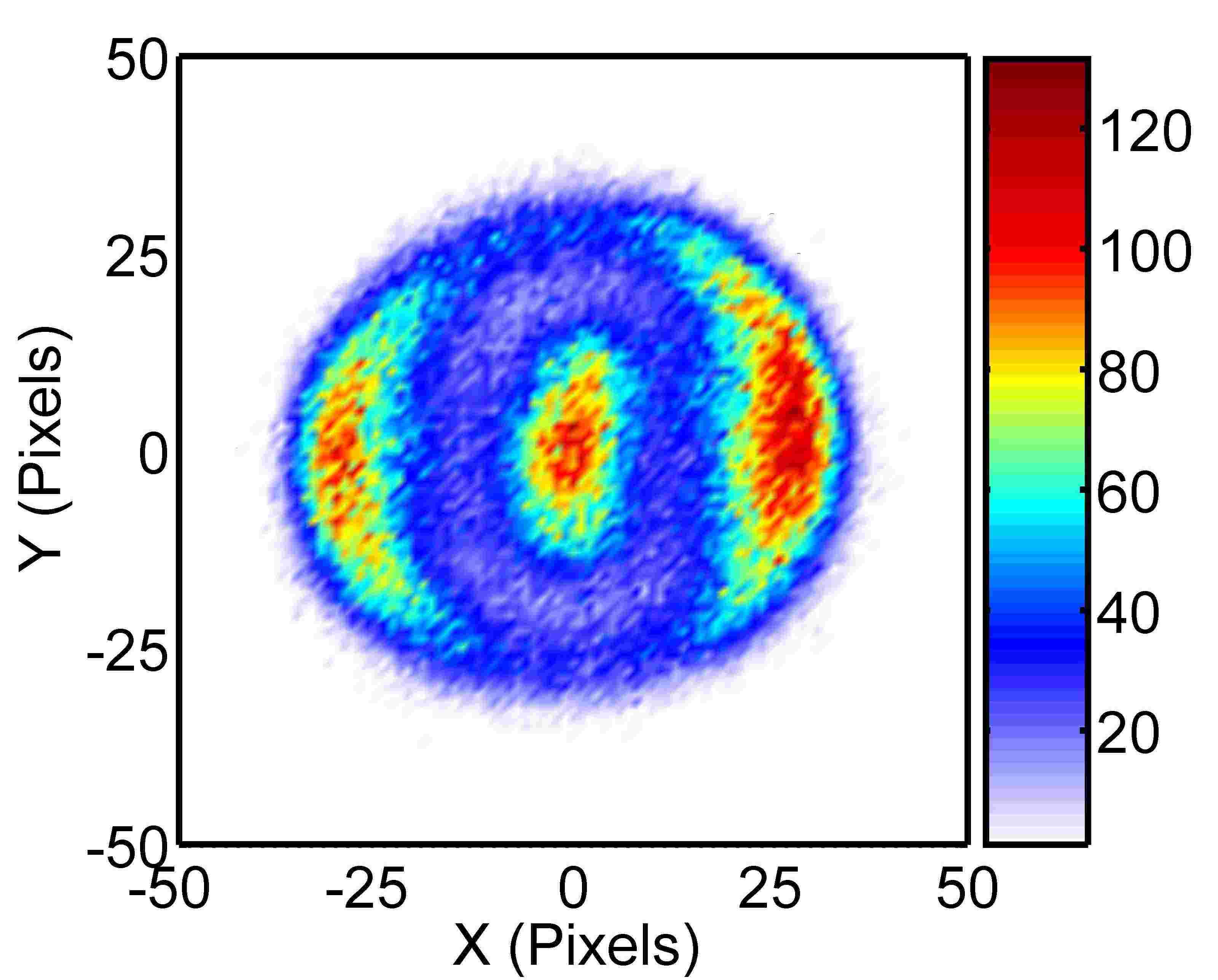}}\\
\subfloat[\ce{H-}/\ce{NH3} at 5.5 eV]{\includegraphics[width=0.3\columnwidth]{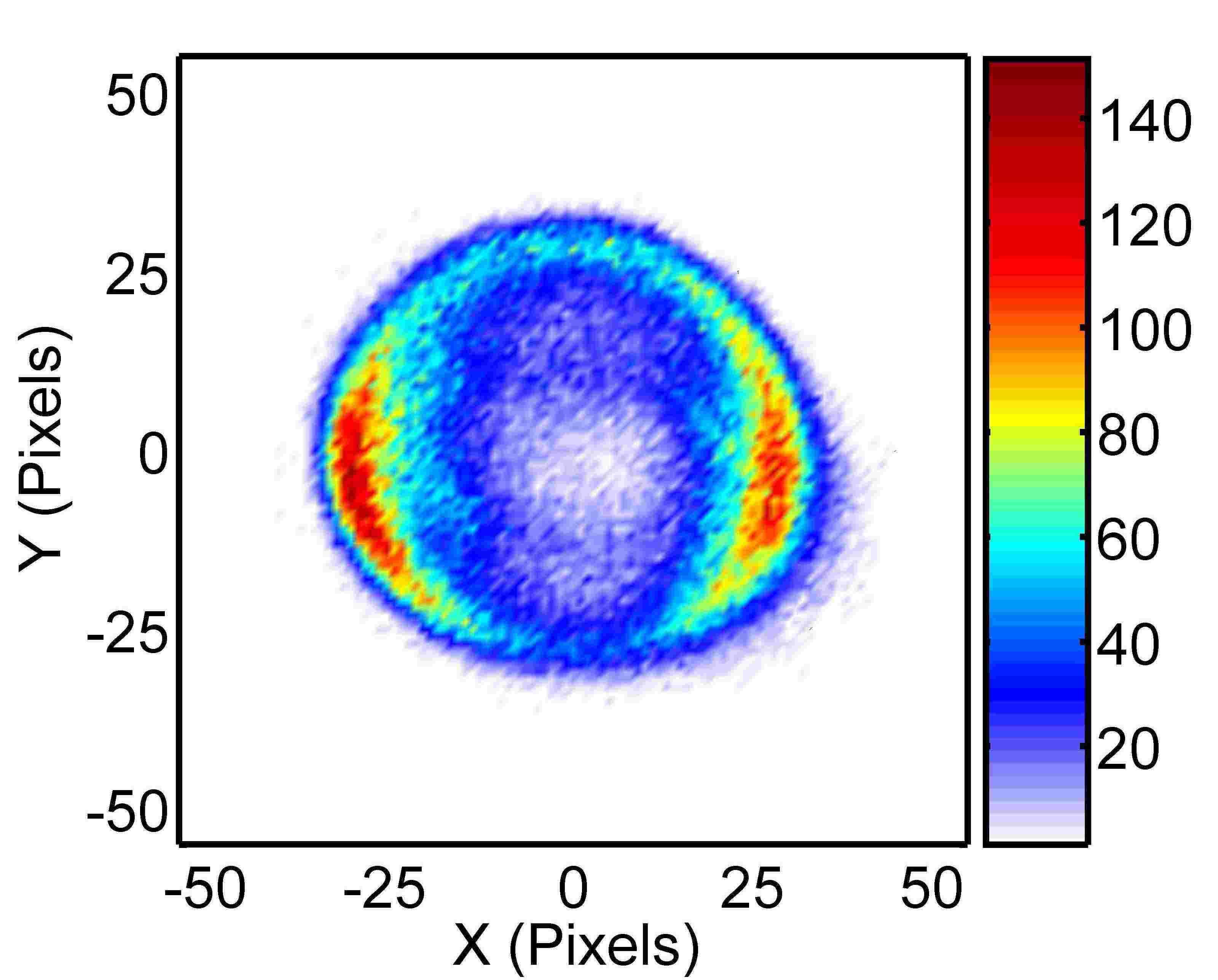}}
\subfloat[KED of \ce{H-}/PA]{\includegraphics[width=0.31\columnwidth]{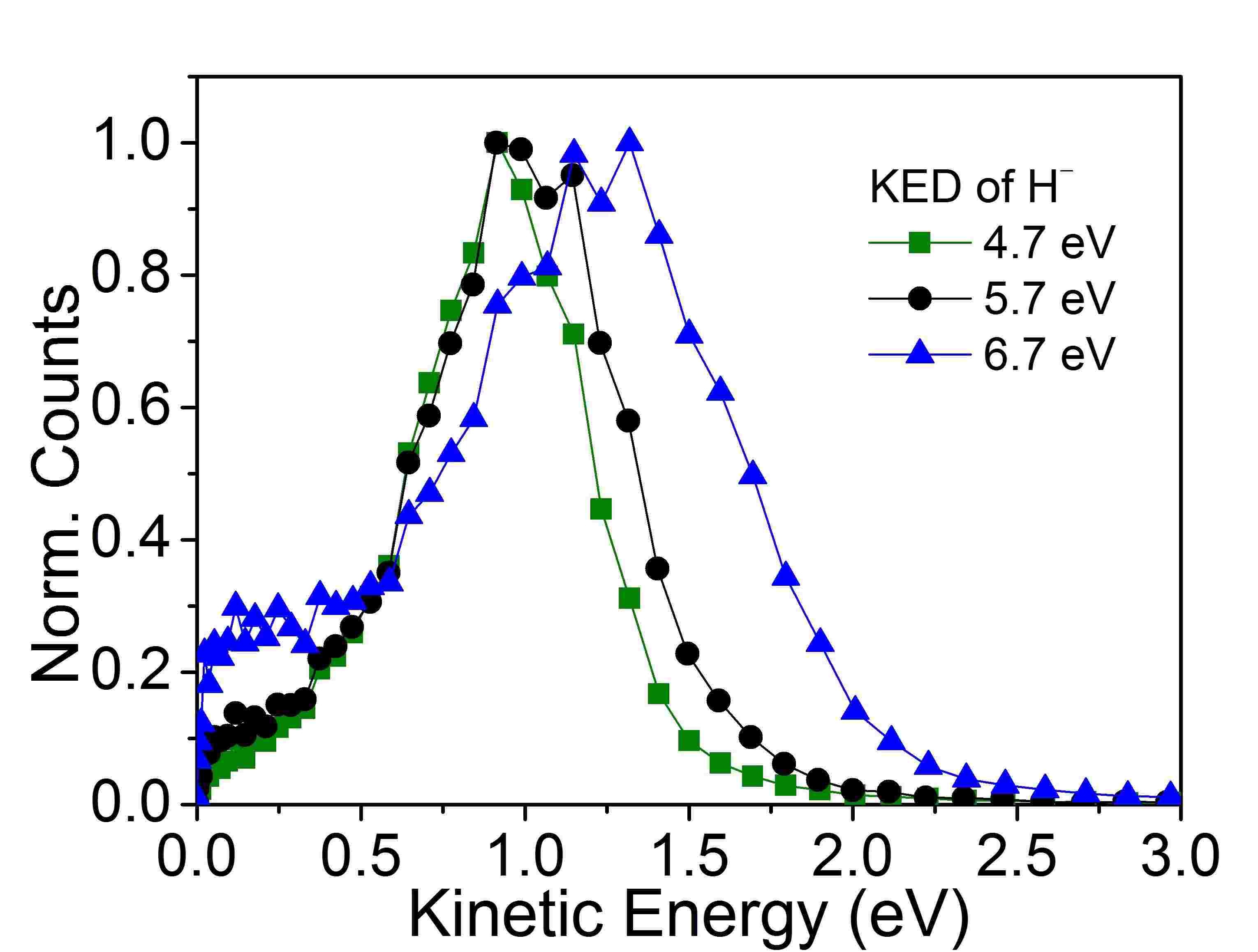}}
\subfloat[Angular distribution plots]{\includegraphics[width=0.31\columnwidth]{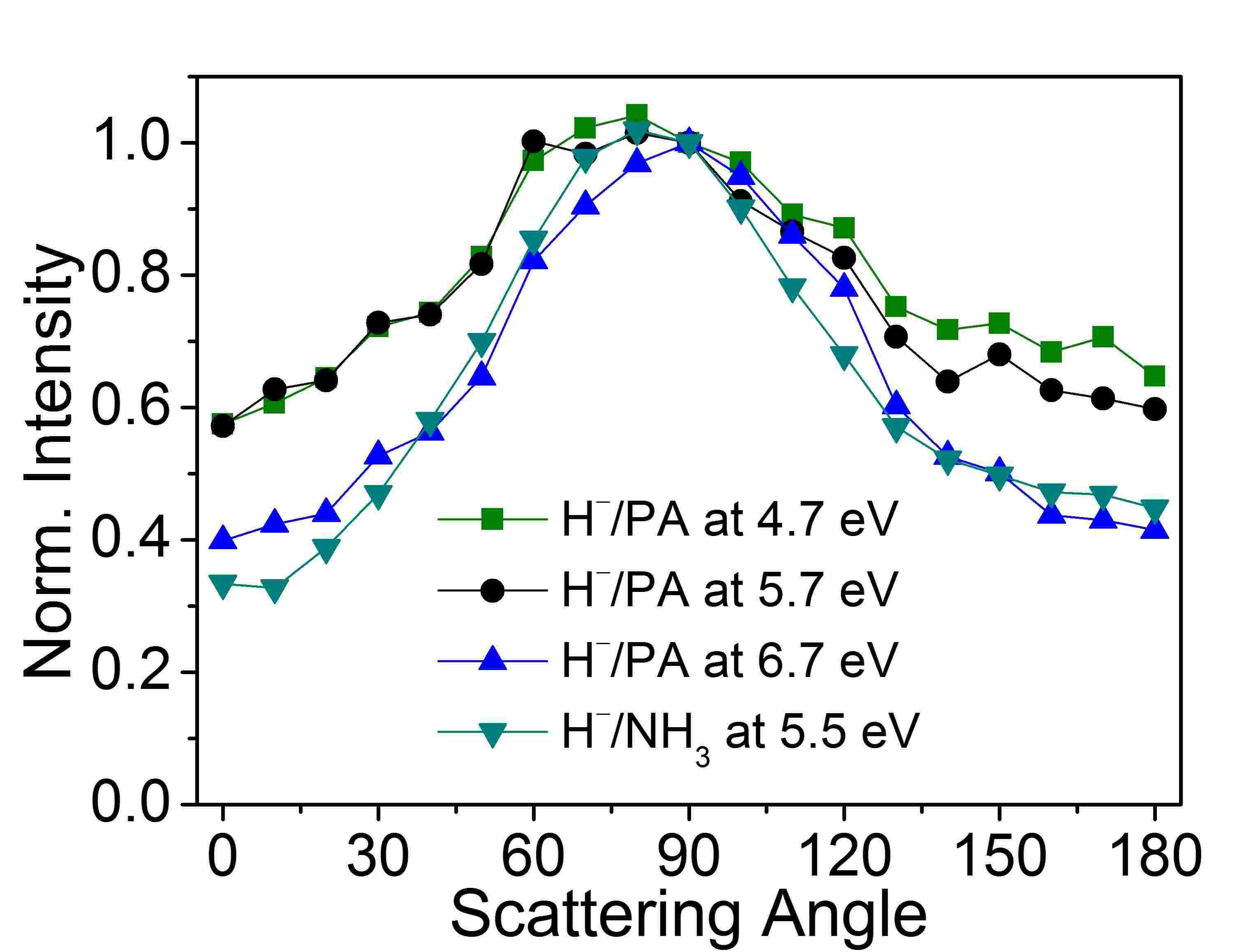}}
\caption{Velocity images of \ce{H-} ions from DEA to Propyl Amine at electron energies (a) 4.5, (b) 5.5 and (c) 6.5 eV along with \ce{H-}/\ce{NH3} at 5.5 eV for comparison. The KE distribution of \ce{H-} ions in (a), (b) and (c) is plotted in (e). Angular distribution curves corresponding to (a), (b), (c) and (d) are plotted in (f).}
\label{fig6.6}
\end{figure}

Figures \ref{fig6.6}(a), (b) and (c) show the velocity images of \ce{H-}  ions scattered at electron energies 4.7 eV, 5.7 eV and 6.7 eV across the resonance along with \ce{H-} from \ce{NH3} at 5.5 eV (Figure \ref{fig6.6}(d)) and are seen to be scattered mostly around the $80^{\circ}$ direction with finite intensity at all other angles. The velocity image at 6.7 eV shows a central blob with very little kinetic energy and becomes dominant at higher electron energies while the outer ring diminishes. The kinetic energy distribution of \ce{H-} ions at the three electron energies are plotted in Figure \ref{fig6.6}(e) and is seen to range from 0 to about 2 eV at 5.7 eV. While data on exact bond dissociation energies of PA is unavailable to deduce the maximum KE of \ce{H-} ions from the NH and CH sites in the molecule, it is seen that the kinetic energy distribution is very similar to \ce{H-} from DEA to \ce{NH3} at 5.5 eV. The resonance at 5.5 eV in Ammonia has been described in detail in paper-III. The KE distribution of \ce{H-} at 5.5 eV (see Figure 5(a) in paper-III) from \ce{NH3} ranges from 0 to 1.8 eV with a peak at 1.2 eV. This similarity in the kinetic energy distribution of \ce{H-} ions from PA and Ammonia suggests \ce{H-} ions emanating from the dissociation of the NH bond in PA. This is further substantiated by the comparison of angular distribution of the \ce{H-} ions at these energies with that of \ce{H-} from Ammonia at 5.5 eV (Figure \ref{fig6.6}(d)). Figure \ref{fig6.6}(f) shows the angular distribution of \ce{H-} ions from PA at 4.7, 5.7 and 6.7 eV along with that of \ce{H-} ions from \ce{NH3} at 5.5 eV. We see the angular distribution curves very similar to each other thereby suggesting that the dissociation dynamics in PA and Ammonia are due to similar resonance behaviour in the 4.5 to 6.5 eV region. Thus, based on this similarity in angular distribution, it is also inferred that the orientation of the NH bond is almost perpendicular to the incoming electron beam. As in the case of Ammonia, it appears that the excitation of the lone pair on the N atom in the amino group is what causes the resonance process about 5.7 eV. Since, the lone pair on N atom has a dominant contribution to the highest occupied molecular orbital - \ce{13a^{'}}, the first resonance process can be attributed to the excitation of the \ce{13a^{'}} electron to the lowest unoccupied orbital \ce{14a^{'}} followed by electron attachment in this first excited state resulting in similar dissociation dynamics as in Ammonia.

\subsubsection{Second resonance process in 8-10 eV region}

\begin{figure}[!htbp]
\centering
\subfloat[\ce{H-}/PA at 7.7 eV]{\includegraphics[width=0.3\columnwidth]{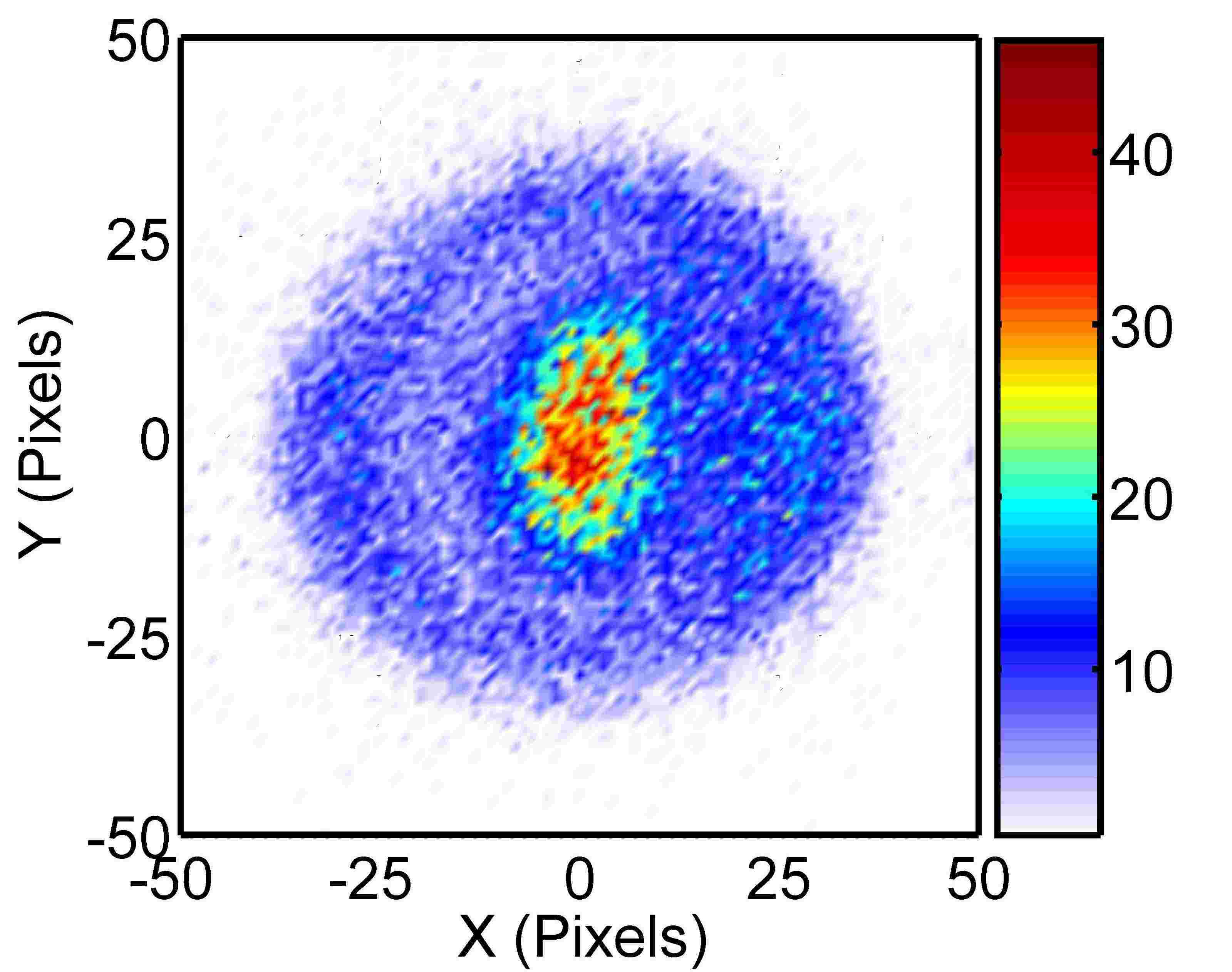}}
\subfloat[\ce{H-}/PA at 8.7 eV]{\includegraphics[width=0.3\columnwidth]{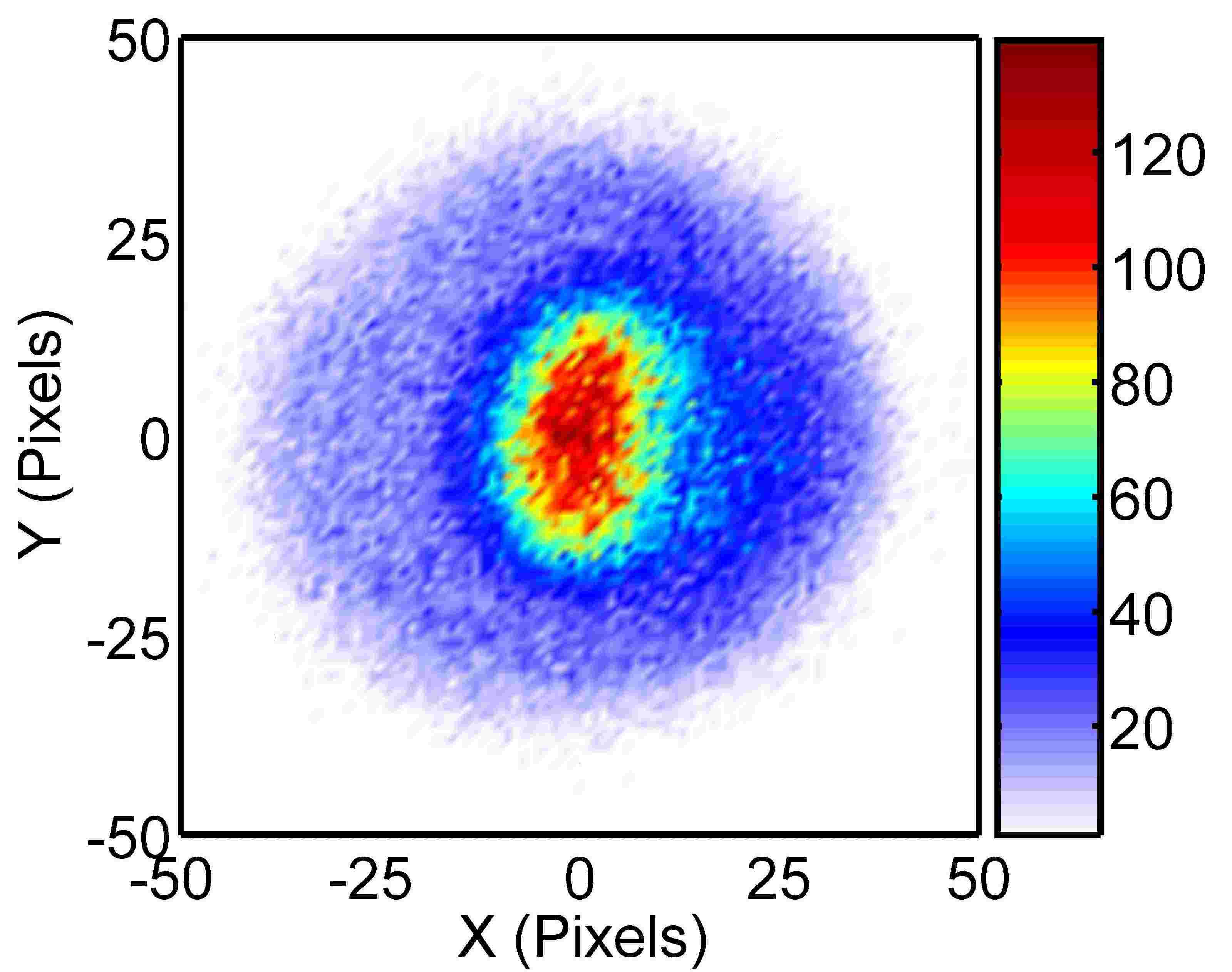}}
\subfloat[\ce{H-}/PA at 9.7 eV]{\includegraphics[width=0.3\columnwidth]{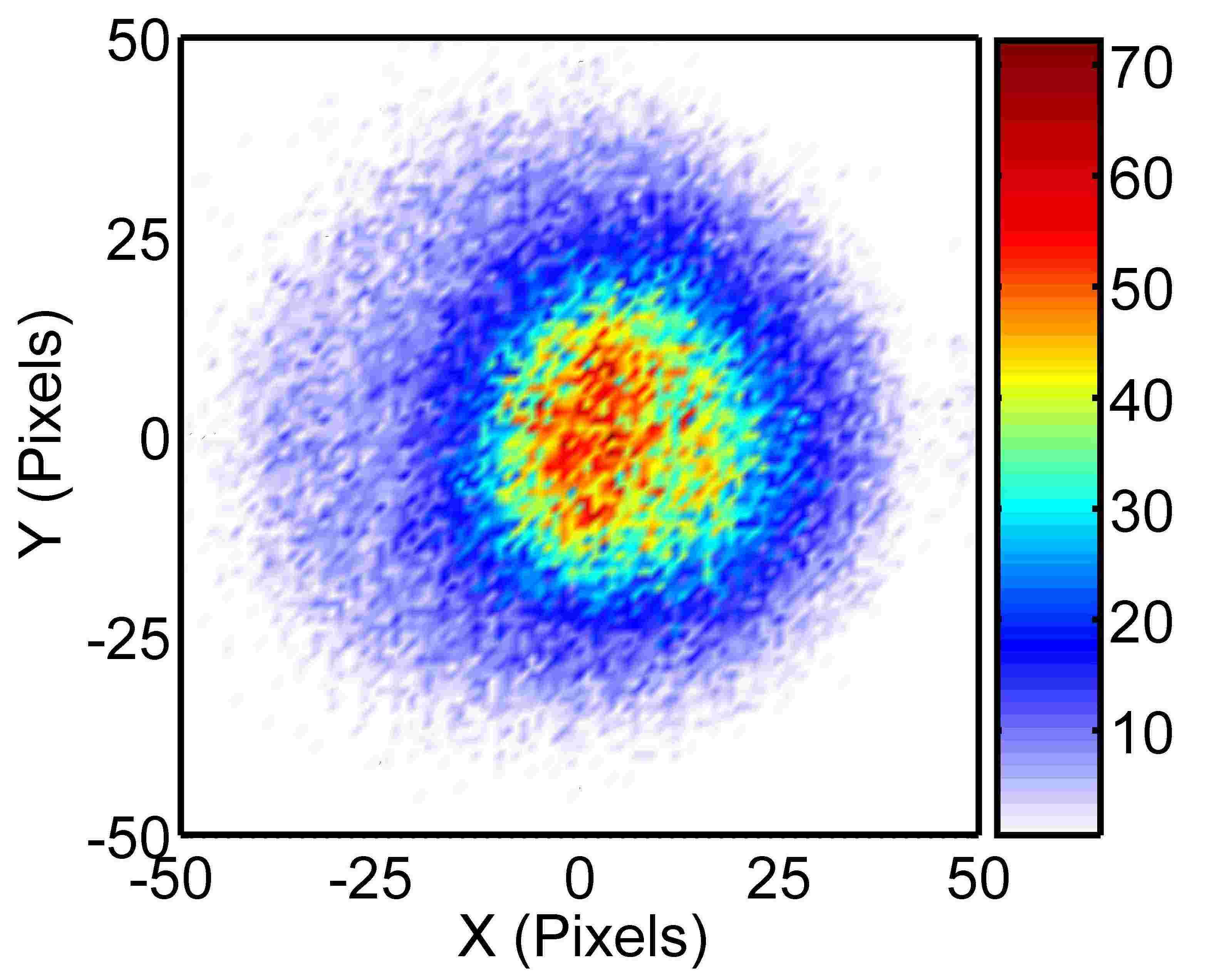}}\\
\subfloat[\ce{H-}/PA at 10.4 eV]{\includegraphics[width=0.3\columnwidth]{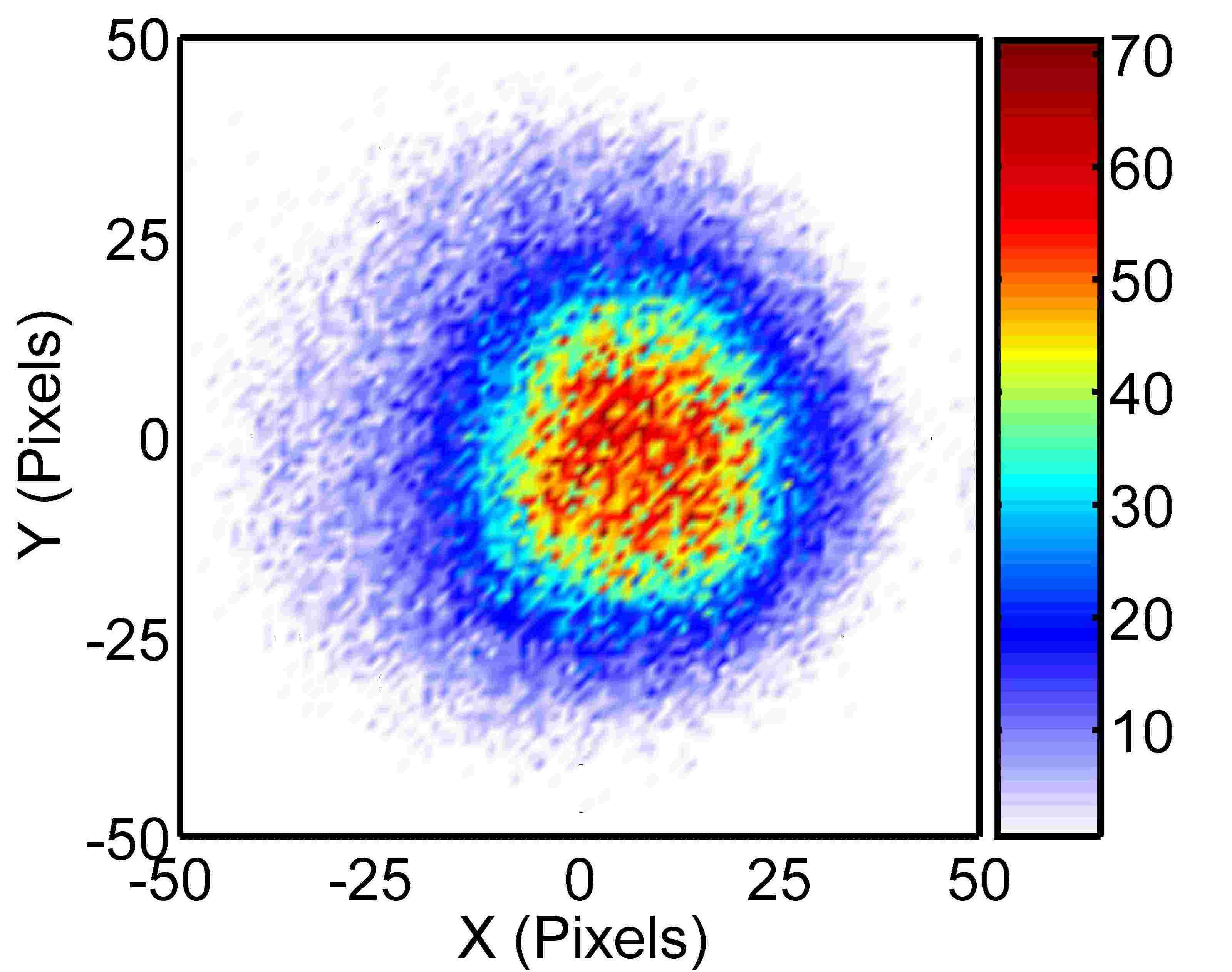}}
\subfloat[\ce{H-}/\ce{CH4} at 10 eV]{\includegraphics[height=4cm,width=5cm]{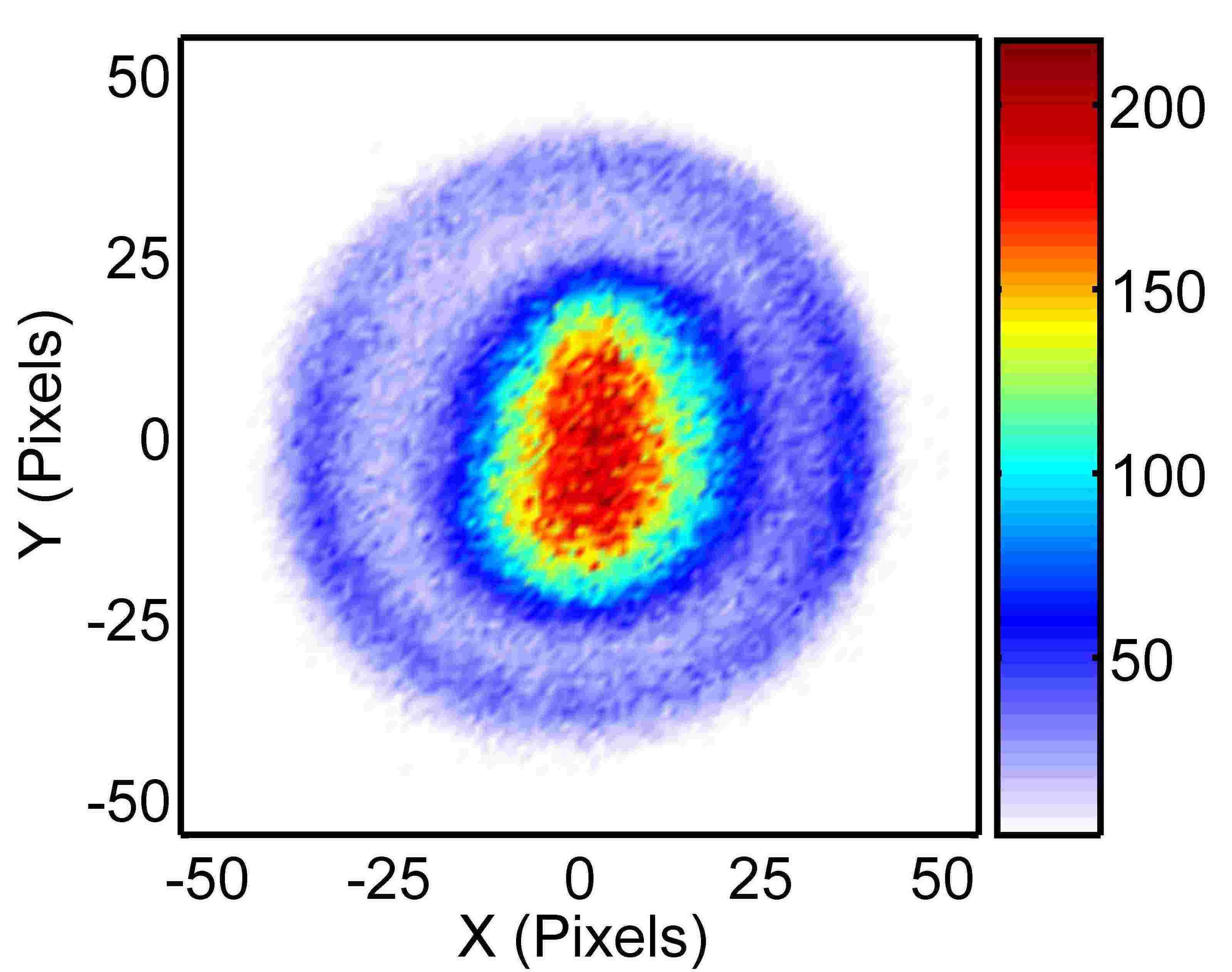}}
\subfloat[KED of \ce{H-} ions]{\includegraphics[height=4cm,width=5cm]{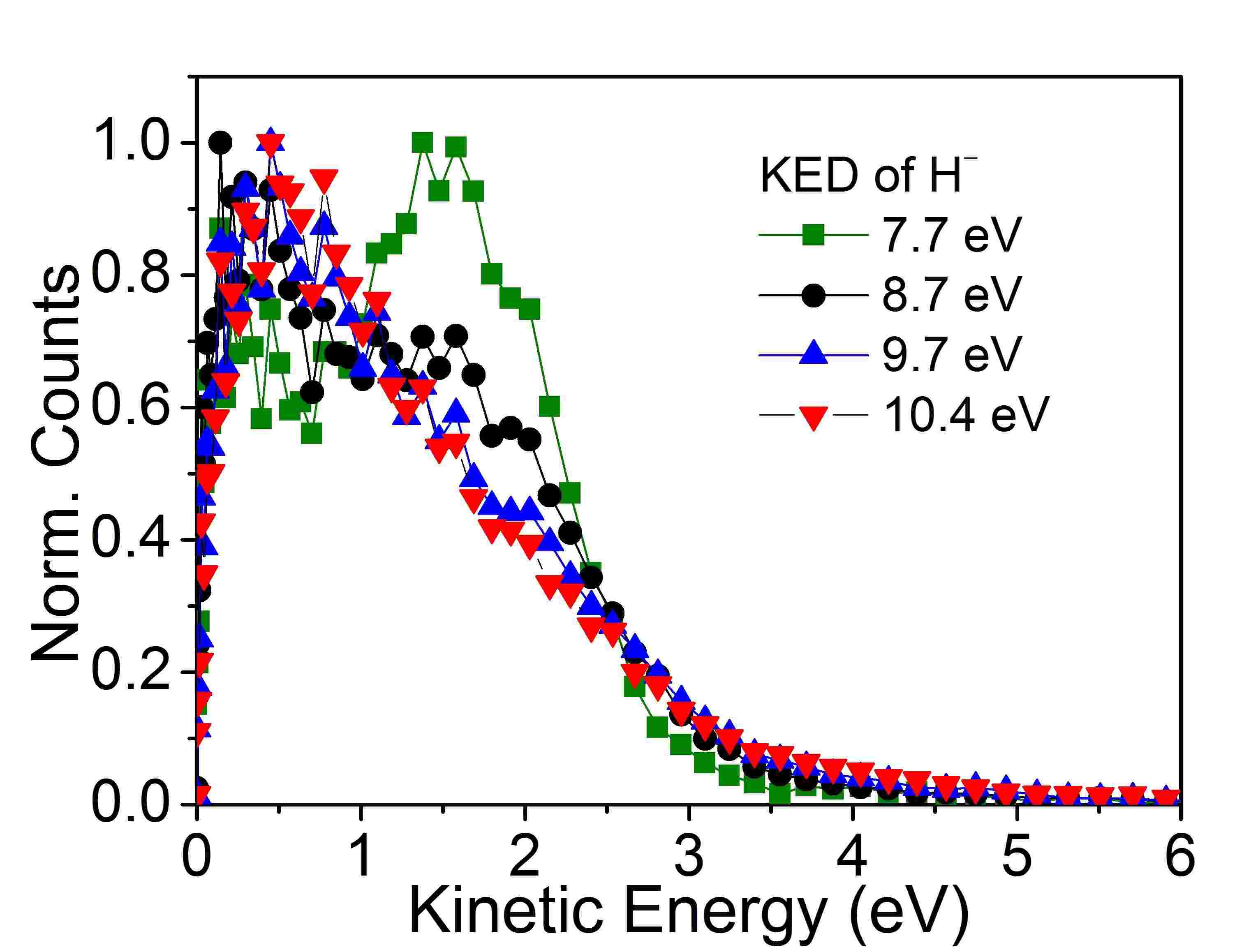}}
\caption{Velocity images of \ce{H-} ions from DEA to PA at electron energies (a) 7.7 eV (b) 8.7 eV (c) 9.7 eV (d) 10.4 eV. For comparison, \ce{H-} from (e) \ce{CH4} at 10 eV are shown. The kinetic energy distribution of \ce{H-} ions in (a), (b), (c) and (d) is plotted in (f).}
\label{fig6.7}
\end{figure}

The velocity images of ce{H-} ions at electron energies 7.7, 8.7 and 9.7 eV are shown in Figures \ref{fig6.7}(a) - (c) along with the velocity image of \ce{H-} from \ce{CH4} at 10 eV (d). These images are characterized by a central blob along with a faint outer ring that is isotropic. The kinetic energy distribution of these ions is shown in Figure \ref{fig6.7}(e) where the maximum KE observed is lower than 4 eV and the most probable energies are in the range 0 to 1.5 eV. We do not have exact bond dissociation energy data on PA to determine the kinetic energies. It is possible that the central blob observed could be emanating from a possible three body breakup scheme. Assuming average bond dissociation energy of 4 eV, the dissociation of either CH or NH bonds in the molecule is expected to give a maximum KE close to 5 eV. It is also possible that the excess energy goes into the excitation of the neutral fragment causing further dissociation. In such a case, lower values of the kinetic energy are observed. An instantaneous 3 body breakup can also account for the low KE energies. Thus, the KE plot in Figure \ref{fig6.7}(e) suggests occurrence of a three body fragmentation along with two body breakup process if any. 

Having argued in favour of a three body fragmentation scheme, it appears that the propyl group is fragmented as against the amino group. Measurements on ammonia at similar energies show \ce{H-} ions produced with high kinetic energies from a two body breakup channel (\ce{H- + NH2^{*}}) and with a strong backward scattering. Whereas, dissociation of Methane negative ion close to 10 eV occurs via a two body as well as a three body channel. While in Methane, this central blob resolves with increase in electron energy giving a forward-backward distribution, such a feature is not seen in PA. The fact that the central blob doesn't resolve into a characteristic anisotropic distribution could be because of strong internal excitation of the propyl group where more vibrational modes are available for soaking up the excess energy. Thus, a particular dissociation limit may not be favoured over the many available ones. Alternatively, the dissociation could be a two step process where the propyl group attaches an electron and is in an excited state which then undergoes dissociation again giving rise to low kinetic energies of \ce{H-} and also the isotropic distribution. Thus, it is inferred that electron attachment to PA in the 8-10 eV region arises from fragmentation of the CH sites with isotropic distribution of the scattered \ce{H-} ions.  While the electron attachment is understood to be from the CH site, the isotropic distribution of ions doesn't give information on the bond orientation at the instant of electron attachment. Strong internal excitations could change the geometry of the molecular anion prior to dissociation and this renders it difficult to infer the bond orientation.

In conclusion, the measurements across the two resonances in PA indicate bond orientation specific dissociation dynamics from the first resonance process peaking at 5.5 eV akin to that in Ammonia as deduced from the close similarity in the velocity images. While the data from second resonance process around 9 eV appear to have features similar to what was seen methane. The very low kinetic energy of the fragments does not allow us to identify signatures of bond orientation specificity in the latter case.

\section{Summary}
\begin{enumerate}
\item \ce{H-} momentum distribution from Formic Acid and Propyl Amine are similar to precursor molecules (water and methane) of the constituent functional groups.

\item Dissociation dynamics independent of overall symmetry of the molecule - Local symmetry of functional group / bond orientation determining factor.

\item Functional group dependence manifests in angular distribution too as seen above.
			\begin{itemize}
				\item OH bond close to 7 eV
				\item CH bond close to 10 eV 
				\item NH bond close to 5.5 eV 
			\end{itemize}

\item Bond orientation rather than the orientation of the whole molecule significant in determining the electron attachment and the angular distribution even in molecule of lower symmetry.
\end{enumerate}

\end{document}